\def\year{2020}\relax
\documentclass[letterpaper]{article} 
\usepackage{aaai20}  
\usepackage{times}  
\usepackage{helvet} 
\usepackage{courier}  
\usepackage[hyphens]{url}  
\usepackage{graphicx} 
\usepackage{graphicx}  
\usepackage{subfigure}
\usepackage{multirow}
\usepackage{xcolor}

\title{Mi YouTube es Su YouTube? \\Analyzing the Cultures using YouTube Thumbnails of Popular Videos
}
 \author{Songyang Zhang
 \quad  Tolga Aktas
 \quad Jiebo Luo\\
 University of Rochester\\
 {\tt\small szhang83@ur.rochester.edu,taktas@u.rochester.edu,jluo@cs.rochester.edu}}

\begin{document}

\maketitle

\begin{abstract}
YouTube, a world-famous video sharing website, maintains a list of the top trending videos on the platform. Due to its huge amount of users, it enables researchers to understand people’s preference by analyzing the trending videos. Trending videos vary from country to country. By analyzing such differences and changes, we can tell how users’ preferences differ over locations.
Previous work focuses on analyzing such culture preferences from videos' metadata, while the culture information hidden within the visual content has not been discovered.
In this study, we explore culture preferences among countries using the thumbnails of YouTube trending videos.
We first process the thumbnail images of the videos using  object detectors.
The collected object information is then used for various statistical analysis. 
In particular, we examine the data from three perspectives: geographical locations, video genres and users' reactions.
Experimental results indicate that the users from similar cultures shares interests in watching similar videos on YouTube.
Our study demonstrates that discovering the culture preference through the thumbnails can be an effective mechanism for video social media analysis.

\end{abstract}

\section{Introduction}
YouTube has been a major social media platform where people share and watch the moments of each other's lives and opinions. Viewers and content makers spend a significant amount of their times to produce and consume content. Besides its prominence in recording a considerable portion of the daily lives, YouTube platform generates massive amounts of data and metadata for each video, allowing researchers to gain an understanding of user preferences and reactions.
YouTube offers a great diversity for the types of videos ranging from news and interviews, to video blogging and friendly conversations. The video content along with the meta-information on every video provides rich information on the interactions of the objects, actions and people, how the audience reacts to such interactions, and how these interactions and reactions vary across cultures. 

Some previous works~\cite{baek2015relationship,park2016data,park2017cultural} focus on video social media analysis using the collected YouTube video metadata. 
However, all these approaches so far ignore the rich information hidden within the video visual content.
Since thumbnails and titles are the two primary sources of information that influence users to watch a video,
understanding the relationship between the thumbnails and the user reaction can shed a light on the underlying cultural preferences in the user's decision making.


In our paper, we focus on exploring the culture preferences hidden within the thumbnails of YouTube trending videos.
In more details, we first detect objects from the thumbnails using a pre-trained object detection framework.
The objects that appear in the thumbnails can be used to  understand the cultural preferences between the viewers in different countries.
This visual information enables us to compare cultures at  a microscopic level.
For example, different countries may be interested in different types of food. Simply relying on the video titles, like "The Best Food in the WORLD," we cannot tell the user preferences from these words without seeing what the food looks like.
Therefore, if we can utilize  the video's object content and explore their relationship with the YouTube metadata, we hope to derive certain new conclusions on the preferences among different cultures. 


Our study analyzes the culture preferences from three perspectives.
First, we analyze the object distribution in thumbnails among different countries.
In more details, we demonstrate that the object distribution vary from country to country.
We also divide all the object labels into several categories (sport, transportation \& traffic, food, animal,  etc.).
We demonstrate that the object distribution among countries varies from category to category.
Besides, we also investigate how the object distribution are associated with some social indexes.

Second, we explore the object distribution in thumbnails among different video genres.
In YouTube, the genre titles are the same in different countries. However, even if the titles are the same, we find that the object distribution in thumbnails in the same genre may still be different among countries. 
Without the visual information, we cannot tell what the difference is.
This finding demonstrates the effectiveness of using visual information in social media analysis.

Third, we investigate how users react to YouTube trending videos in different countries.
In more details, we are interested in how average views, average like per view, and average comment per view in one country differ from others.
Through various analysis, we find that the underlying cultural visual preferences can also influence the user's decision making.


IN summary, we investigate the following three key research questions and contribute findings and insights:

\begin{itemize}
  \item What is the distribution of objects in the thumbnails in different countries?
  \item Is the correlation between objects in thumbnails and video genres the same among different countries?
  \item How do YouTube users react given the object distribution in thumbnails in different countries?
  
\end{itemize}

\section{Related Work}

Recent video studies in the computer vision field focus on understanding visual concept from raw video data without considering the underlying culture factors. 
Meanwhile, video studies in the data mining field have concentrated on the metadata (duration, tag, publish time, etc.) without considering video content.
Our approach bridge the gap between these two research threads.

\subsection{Image based Social Media Analysis}
A number of image-based social media analyses, such as fashion studies~\cite{simo2015neuroaesthetics}, political analysis~\cite{joo2014visual} and architectural style~\cite{doersch2015makes}, have been performed to gain insights into the web images.
Recently, there are also several studies on culture related topics.
Zhou \textit{et al.} (2018)
propose a method to characterize different cities from 2 million geo-tagged photos via performing an attribute analysis between the city and image contents of the city.
You \textit{et al.} (2017) 
look at the problem of discovering spatio-temporal evolution of cultural preferences for lifestyles from anonymized Facebook photographs. 
Khosla \textit{et al.} (2014)
learn the image popularity from the image content and social context.
Redi \textit{et al.} (2016)
coin the term "photo cultures" and analyze how the choices of style varies in photography and architecture across different cultures and locations.
In contrast, our work studies the visual concepts among different countries by analyzing YouTube trending videos, providing a novel perspective for studying the culture preference.

\subsection{Video based Social Media Analysis}
Recently, some approaches focus on video based cultural analysis.
They study the user preferences by collecting YouTube data from various perspectives.
Park \textit{et al.} (2016)
show that the video view duration is positively associated with the video’s view count, the number of likes per view, and the negative sentiment in the comments.
Baek \textit{et al.} (2015)
demonstrate that the Korean pop (or K-pop) music videos trending on YouTube are less related to the culture closeness with Korea.
Park \textit{et al.} (2017)
investigate the popular videos in countries that differ in cultural values, language, gross domestic product, and Internet penetration rate.
Different from these previous works, our study focuses on analyzing the culture preference and the video content. Moreover, our approach is not limited to any specific type of media content or any specific countries. 

\subsection{Video based Video Attractiveness Analysis}
 Some studies emphasize the prediction of the attractiveness from a given video.
This task was first proposed by Liu \textit{et al.} (2009).
They measure the interestingness of the video frames using Flickr images.
Later on, Jiang \textit{et al.} (2013)
collect two interestingness prediction datasets from Flickr and YouTube.
Recently, Chen \textit{et al.} (2018)
tackle the fine-grained video attractiveness prediction problem by utilizing the technology of moment localization with natural language~\cite{gao2017tall,hendricks17iccv}.
In contrast to implicitly predicting video attractiveness from a given video, our goal is to explicitly find what kind of objects in thumbnails can interest the users in different cultures.



\section{Data}

Our dataset is extended from the Trending YouTube Video Statistics dataset~\cite{kaggle} with additionally collected video thumbnails. 
In the original dataset, the metadata (video title, channel title, publish time, tags, views, likes and dislikes, description, comment count and thumbnails link) on daily trending YouTube videos are collected, aggregated from \textit{Nov $2017$} to \textit{Jun $2018$}. Videos come from US, GB, DE, CA, FR, MX, RU, IN, JP and KR (United States, Great Britain, Germany, Canada, France, Mexico, Russia, India, Japan, and South Korea, respectively), with up to 200 listed trending videos per day.
In our extension, we download all video thumbnails through the links provided by the metadata.
After removing the videos with invalid links, video ids or dates, there are approximately $150,000$ videos. Note that only these valid videos are included in our analysis.

\begin{table*}[t]
    \centering
    \begin{tabular}{|c|c|}
    \hline
    Category & Labels\\
    \hline
    \hline
    \multirow{2}{*}{Sport} & \textit{frisbee}, \textit{skis}, \textit{snowboard}, \textit{sports ball}, \textit{kite}, \textit{baseball bat}, \\
    & \textit{baseball glove}, \textit{skateboard}, \textit{surfboard}, \textit{tennis racket} \\
    \hline
    \multirow{2}{*}{Transportation \& Traffic} & \textit{bicycle}, \textit{car}, \textit{motorcycle}, \textit{airplane}, \textit{bus}, \textit{train}, \textit{truck}, \textit{boat}, \\
    & \textit{traffic light}, \textit{fire hydrant}, \textit{stop sign}, \textit{parking meter} \\
    \hline
    \multirow{1}{*}{Food} & \textit{banana}, \textit{apple}, \textit{sandwich}, \textit{orange}, \textit{broccoli}, \textit{carrot}, \textit{hot dog}, \textit{pizza}, \textit{donut}, \textit{cake}\\
    \hline
    \multirow{1}{*}{Animal} & \textit{bird}, \textit{cat}, \textit{dog}, \textit{horse}, \textit{sheep}, \textit{cow}, \textit{elephant}, \textit{bear}, \textit{zebra}, \textit{giraffe}\\
    \hline
    \multirow{2}{*}{Kitchenware} & \textit{dining table}, \textit{microwave}, \textit{oven}, \textit{sink}, \textit{refrigerator}, \textit{knife}, \textit{spoon}, \textit{bottle}, \\ 
    & \textit{wine glass}, \textit{cup}, \textit{bowl}, \textit{fork}, \textit{toaster}\\
    \hline
    \multirow{2}{*}{Household} & \textit{bed}, \textit{tv}, \textit{bench}, \textit{laptop}, \textit{remote}, \textit{toilet}, \textit{potted plant}, \textit{mouse}, \textit{keyboard}, \\
    & \textit{cell phone}, \textit{clock}, \textit{vase}, \textit{scissors}, \textit{teddy bear}, \textit{hair drier}, \textit{toothbrush}, \\
    & \textit{chair}, \textit{coach}, \textit{umbrella}, \textit{vase}, \textit{tie}, \textit{book}\\
    \hline
    Belongings & \textit{suitcase}, \textit{handbag}, \textit{backpack} \\
    \hline
    Person & \textit{person}\\
    \hline
    \end{tabular}
    \caption{Object labels list used in our analysis. All labels in the table are the same as COCO annotations~\cite{lin2014microsoft}. We further divide them into several categories used for the following analysis.}
    \label{tab:object_class}
\end{table*}

\begin{figure*}[t!]
    \centering
    \includegraphics{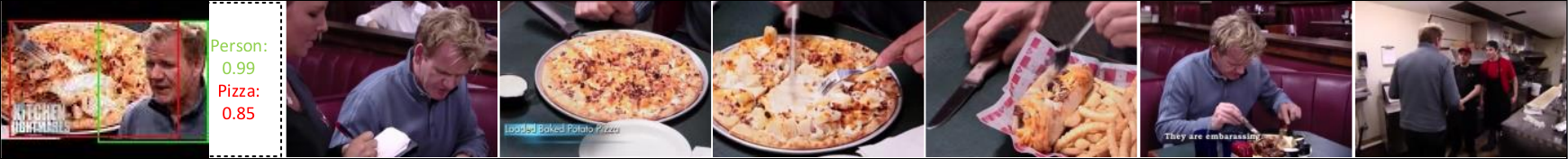}
    \includegraphics{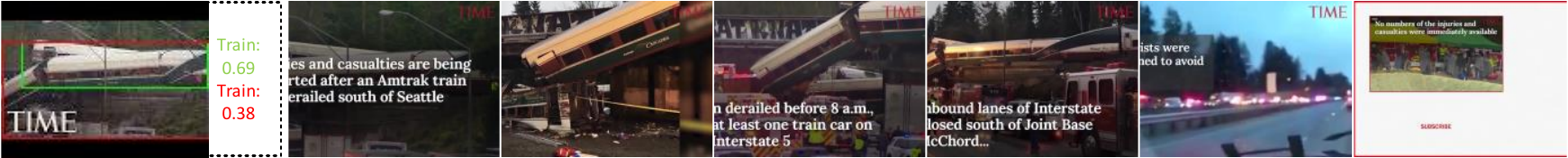}
    \caption{Example thumbnails and their corresponding videos with the top two detected objects, located in the green and red boxes, respectively. The predicted labels and scores are listed on the right side.}
    \label{fig:examples}
\end{figure*}

\section{Methodology}

We are interested in analyzing the YouTube video thumbnails between different countries. 
To quantify the culture differences, we first need to identify object labels from thumbnails.
Then we conduct several statistical analyses based on the detected objects and metadata.

To predict the objects occurred in thumbnails, we use Faster-RCNN~\cite{ren2015faster}, which has shown the state-of-the-art performance on the object detection task.
Specifically, we adopt a Faster-RCNN with ResNet50 as the  backbone and pre-trained it on MS-COCO dataset~\cite{lin2014microsoft}. We follow the details and hyper-parameters specified in the original paper. See~\cite{ren2015faster} for the full details (\url{https://pytorch.org/tutorials/intermediate/torchvision_tutorial.html}).
In the following analysis, we only consider the existence of the objects in the thumbnails.
If an object exists in the thumbnails, we count it as $1$, otherwise, we count it as $0$. It is noteworthy that although one could analyze the entire videos for object recognition, the sheer amount of video data makes it computationally prohibitive. This is why we chose to  analyze only the video thumbnails in this study.

When we introduce a discussion of culture, our approach focuses on the common objects that attract the users' attention. Specifically, we concentrate on the different distribution of the objects in thumbnails among different countries and video genres.

For analysis, we first perform statistical analyses to determine the distribution of the objects among different countries and genres.
Comparisons between countries, and between country and  social indexes are also conducted.
We then investigate the object distribution among different genres.
We also show how much the object distribution of one genre may vary among different countries.
Finally, we explore the relationship between object distribution and views, likes per view and comments per view, and investigate the similarities between countries in terms of these three factors.

\section{Results}

\begin{figure*}[t!]
\centering
\begin{tabular}{ccc}
\includegraphics[width=5.6cm]{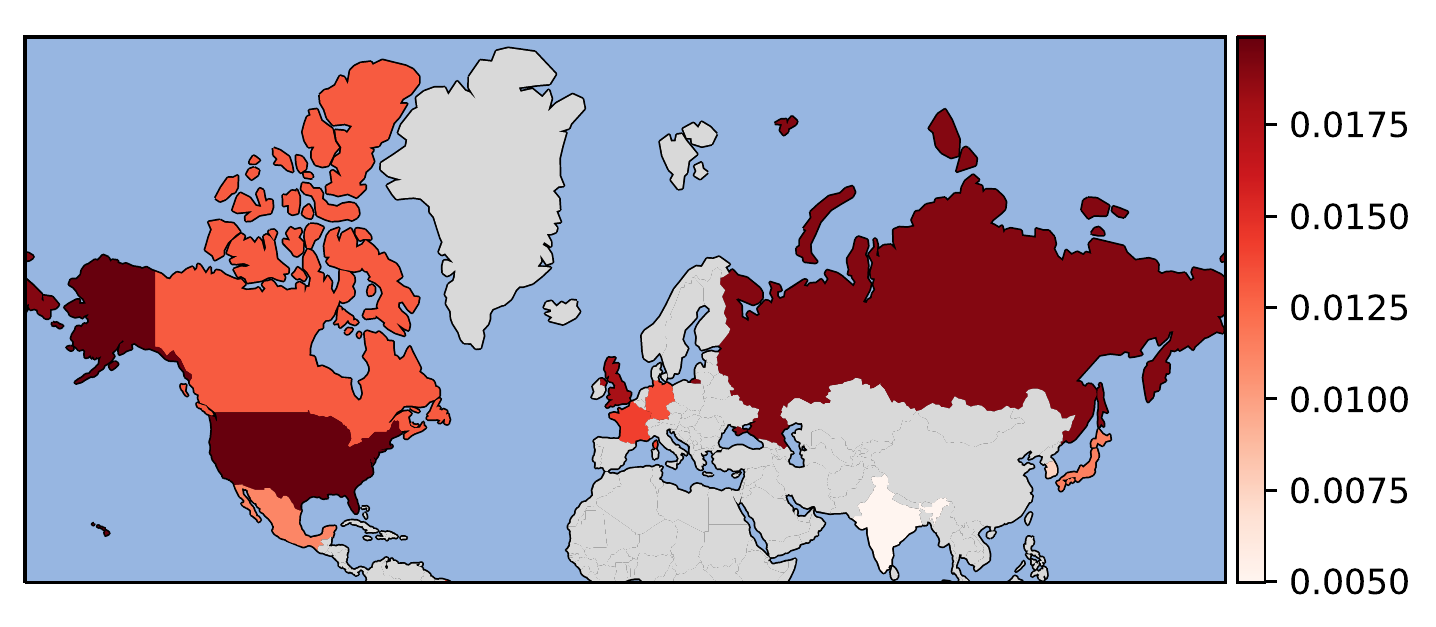} &
\includegraphics[width=5.6cm]{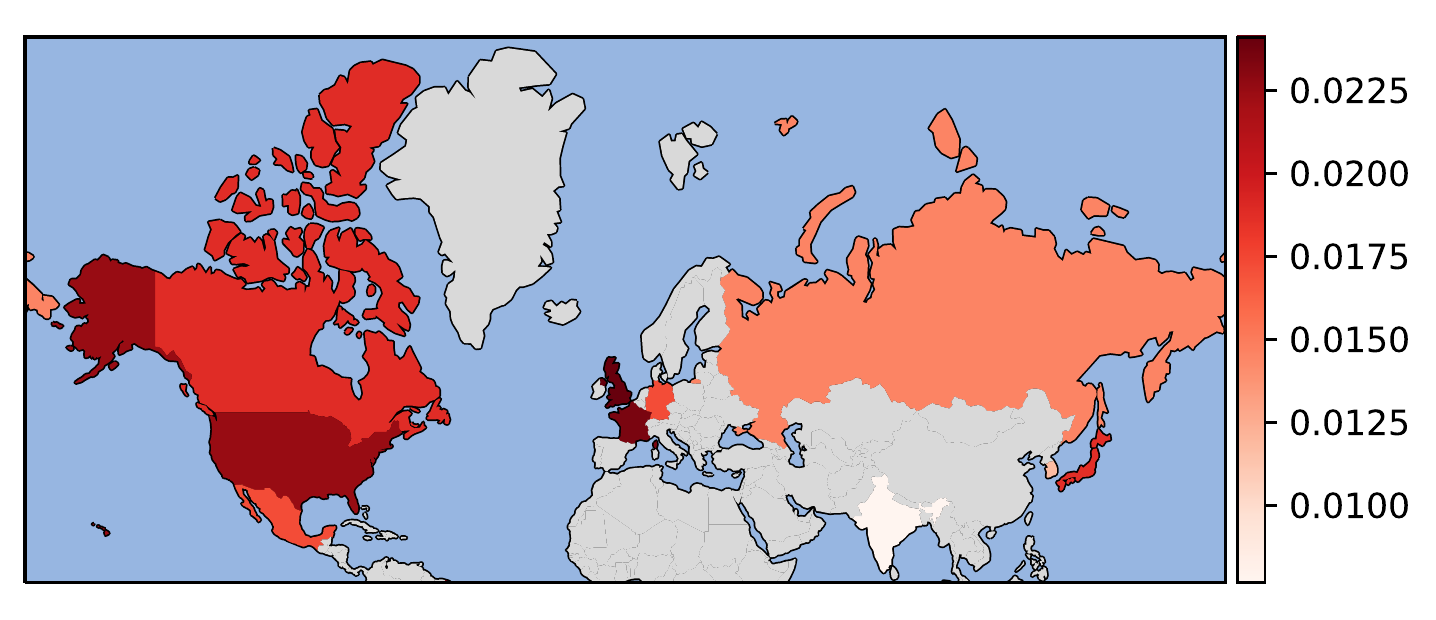} &
\includegraphics[width=5.6cm]{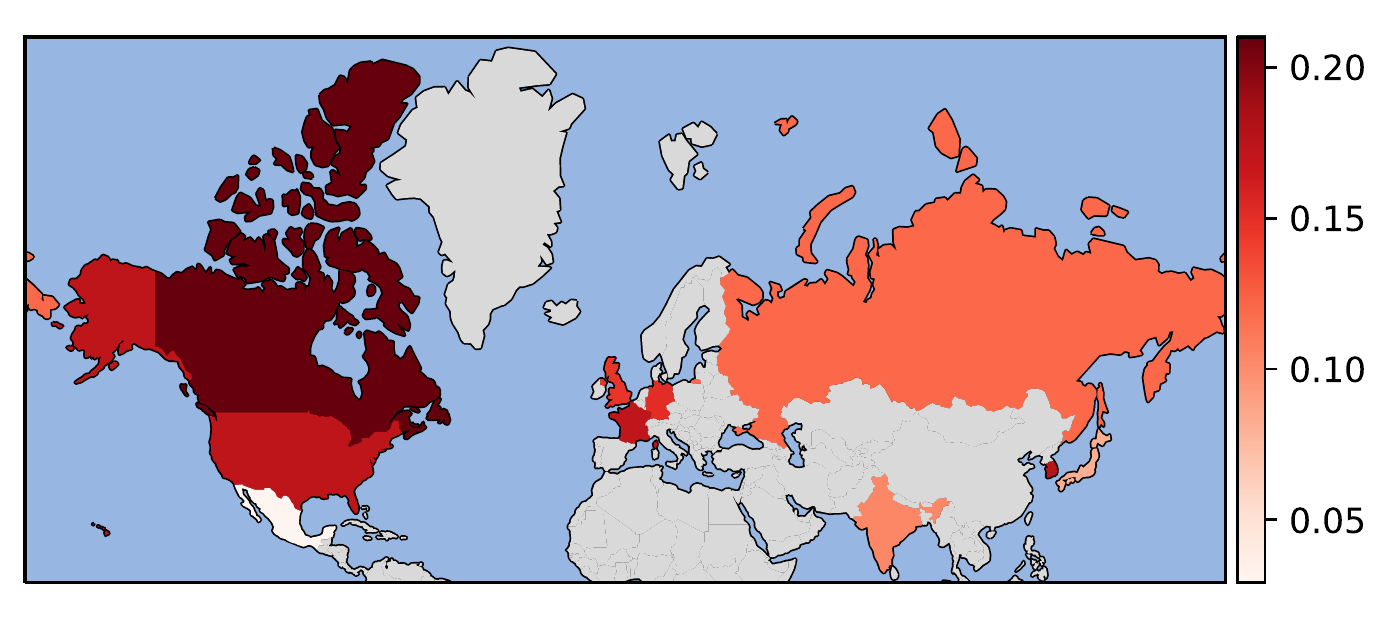} \\
(a) Airplane & (b) Bicycle & (c) Bus \\ 
\includegraphics[width=5.6cm]{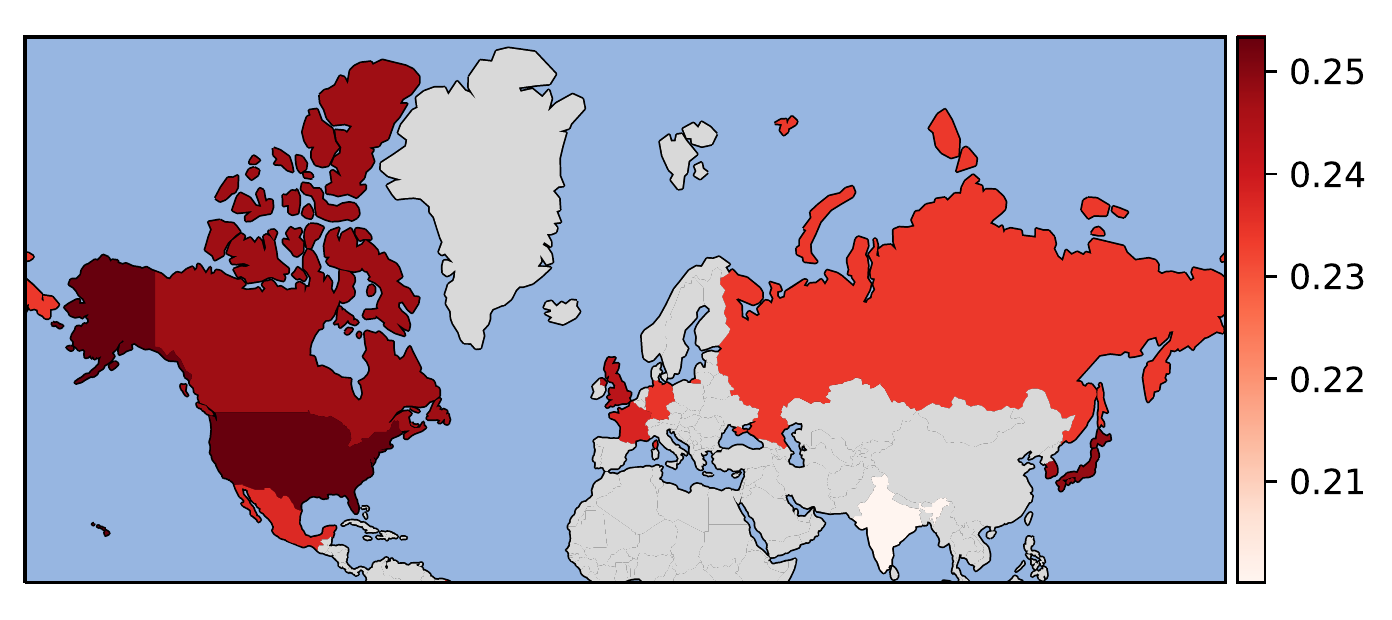} &
\includegraphics[width=5.6cm]{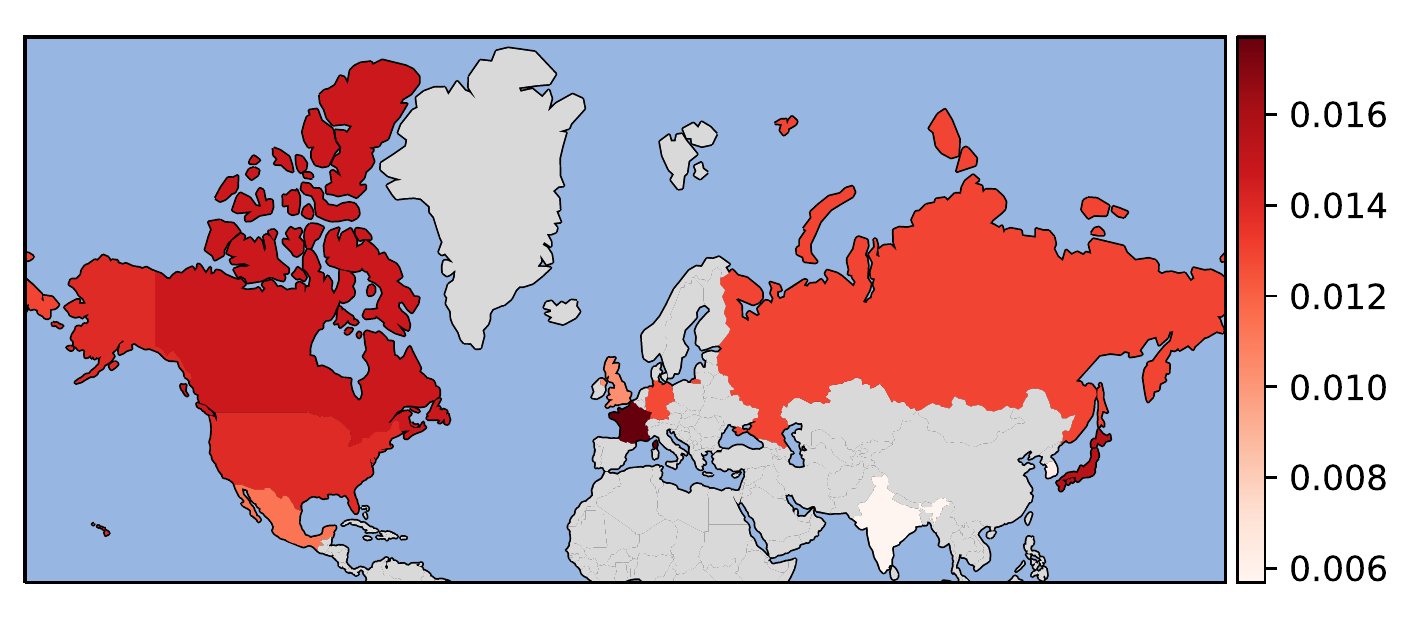} &
\includegraphics[width=5.6cm]{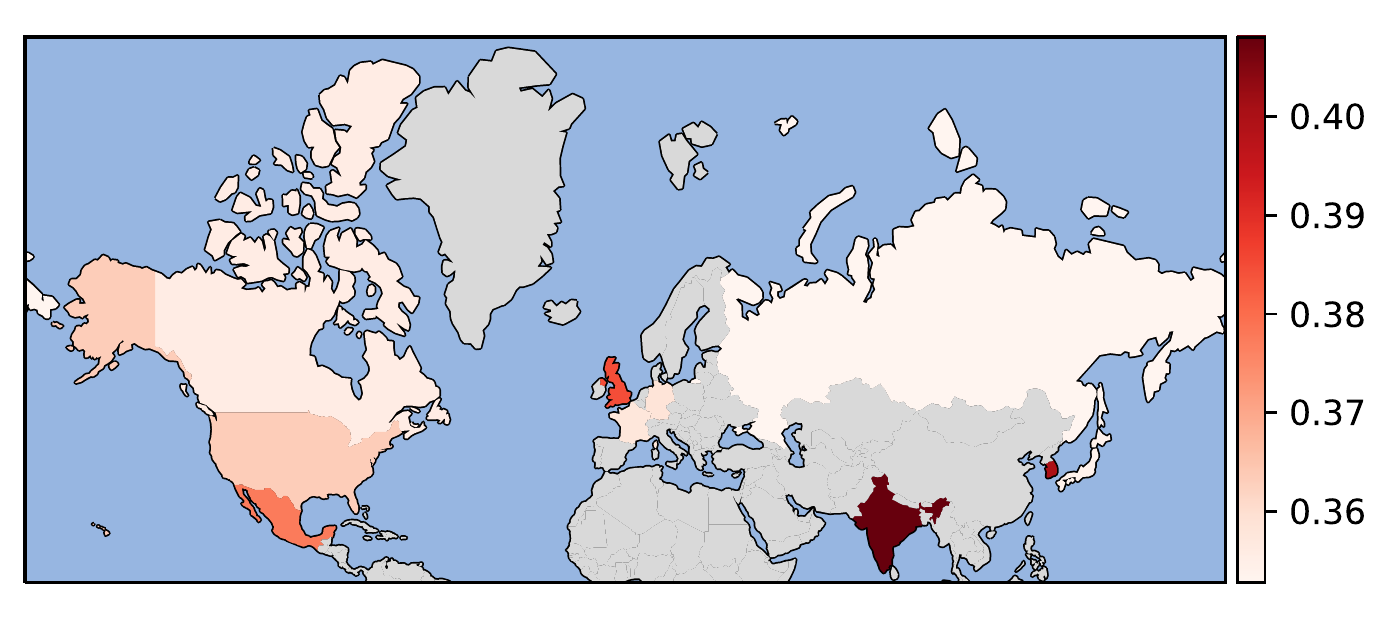} \\
(d) Car & (e) Motorcycle & (f) Train \\ 
\includegraphics[width=5.6cm]{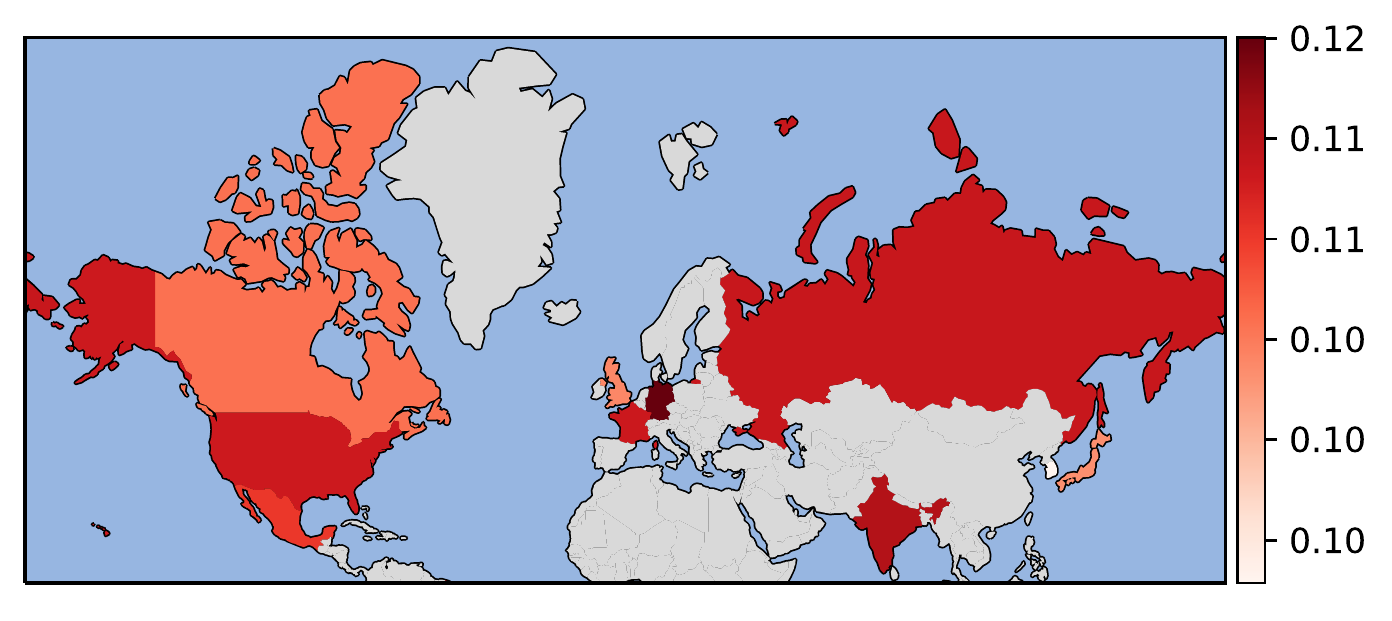} & \includegraphics[width=5.6cm]{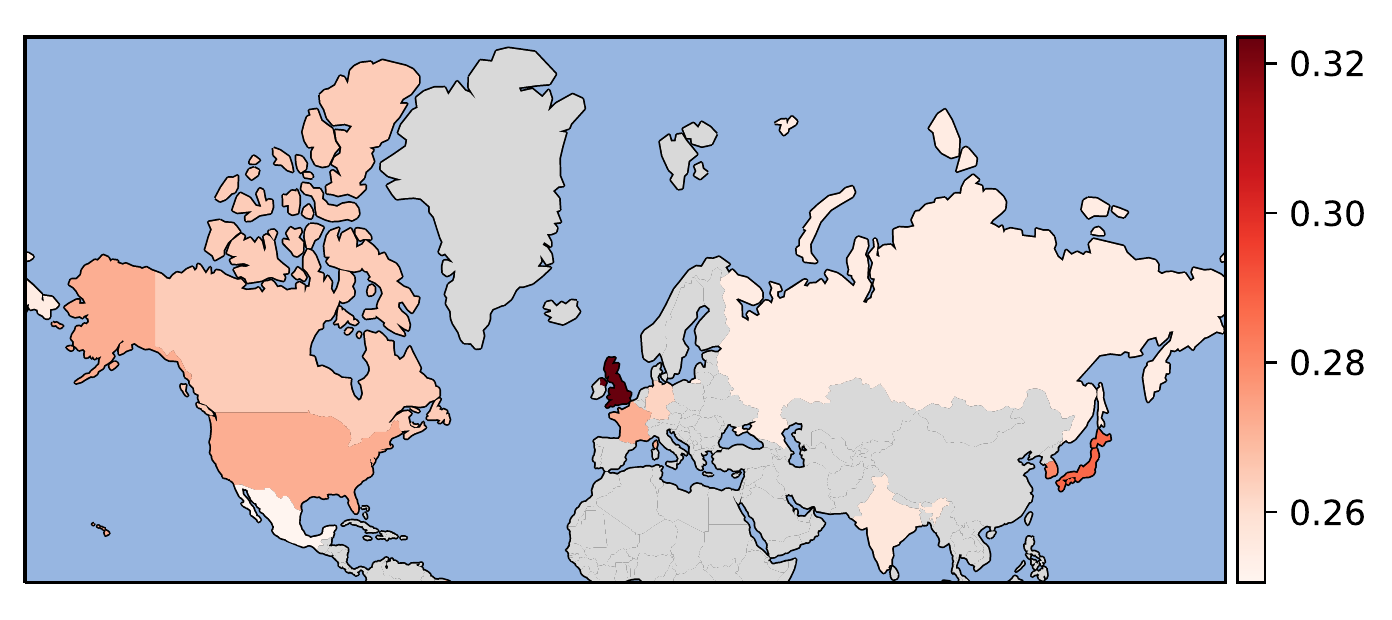} &
\includegraphics[width=5.6cm]{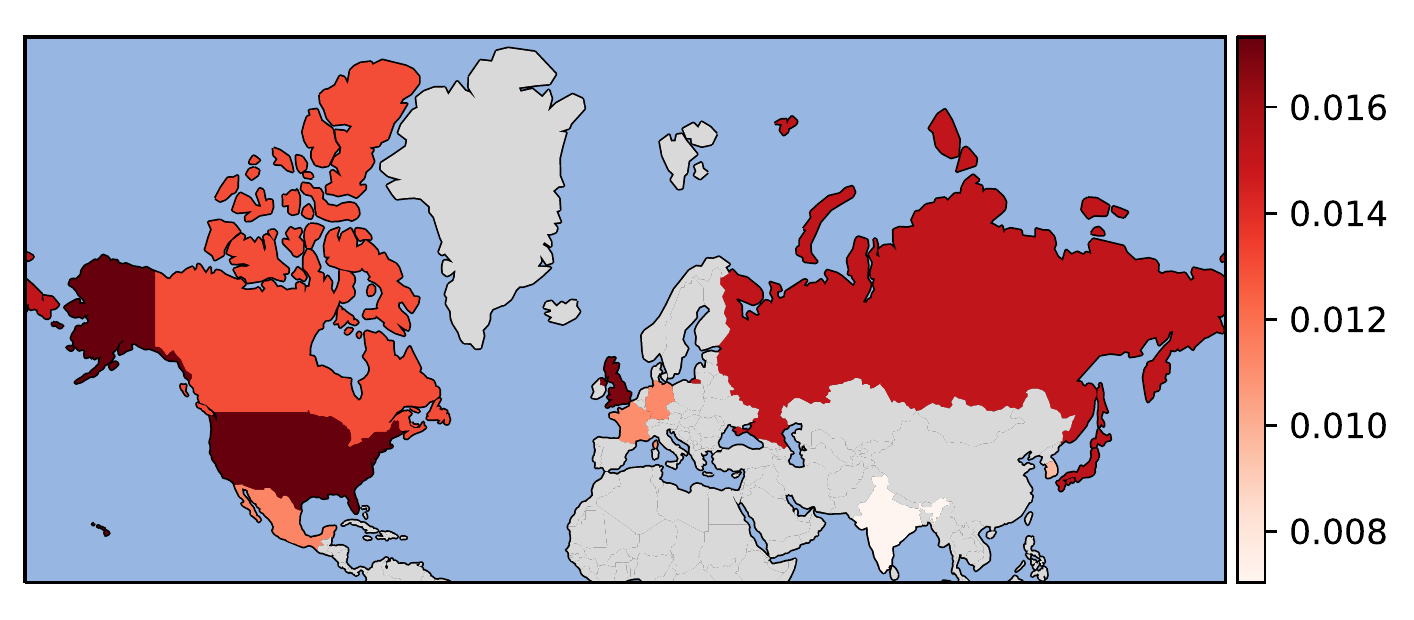} \\
(g) Boat & (h) Baseball bat or glove & (i) Skis \\ 
\includegraphics[width=5.6cm]{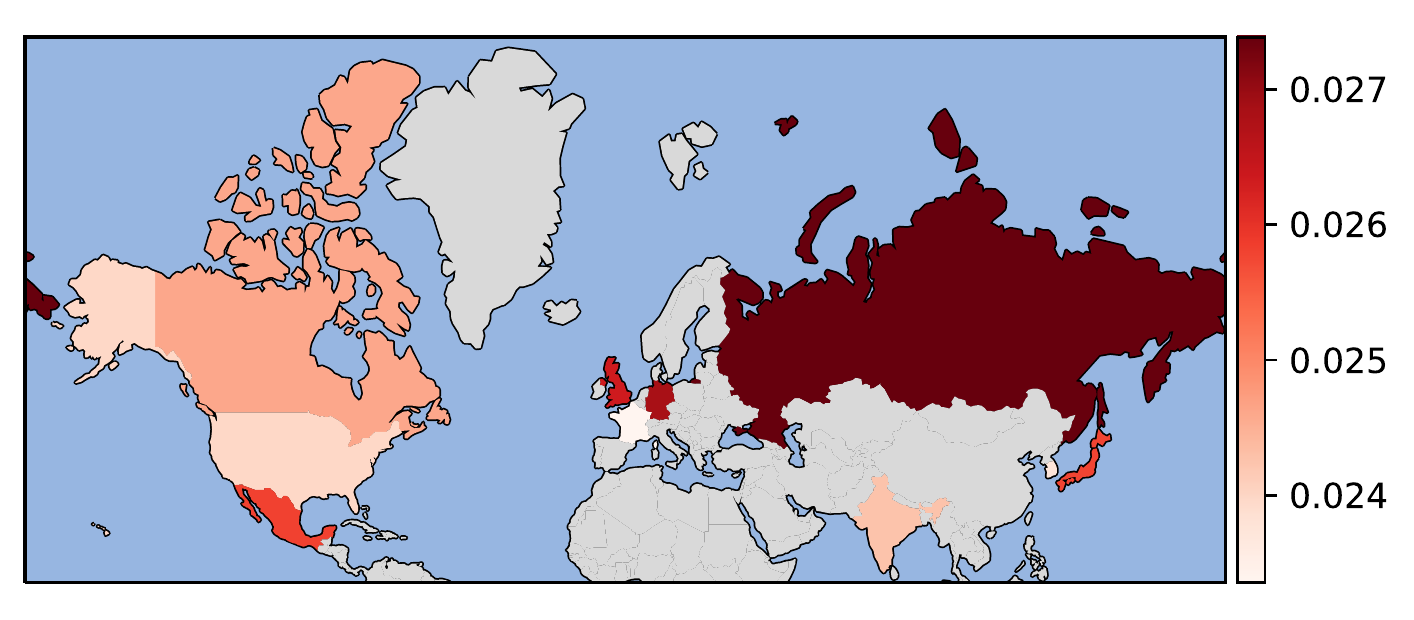} &
\includegraphics[width=5.6cm]{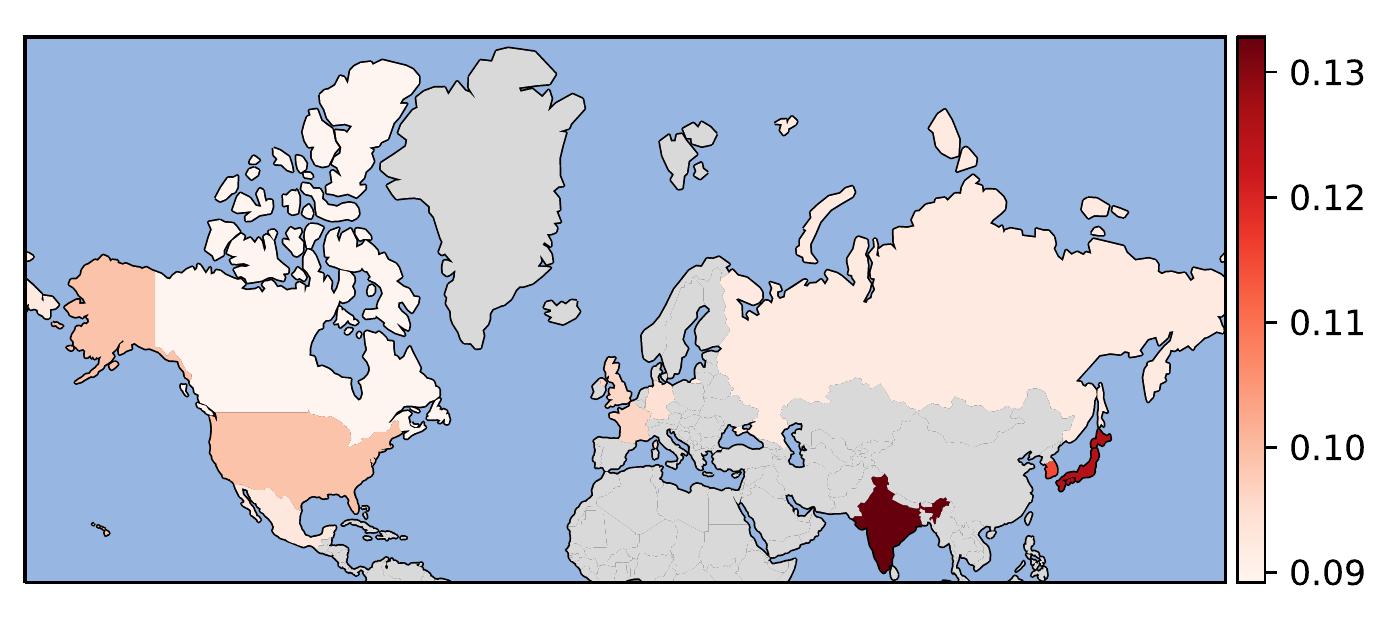} &
\includegraphics[width=5.6cm]{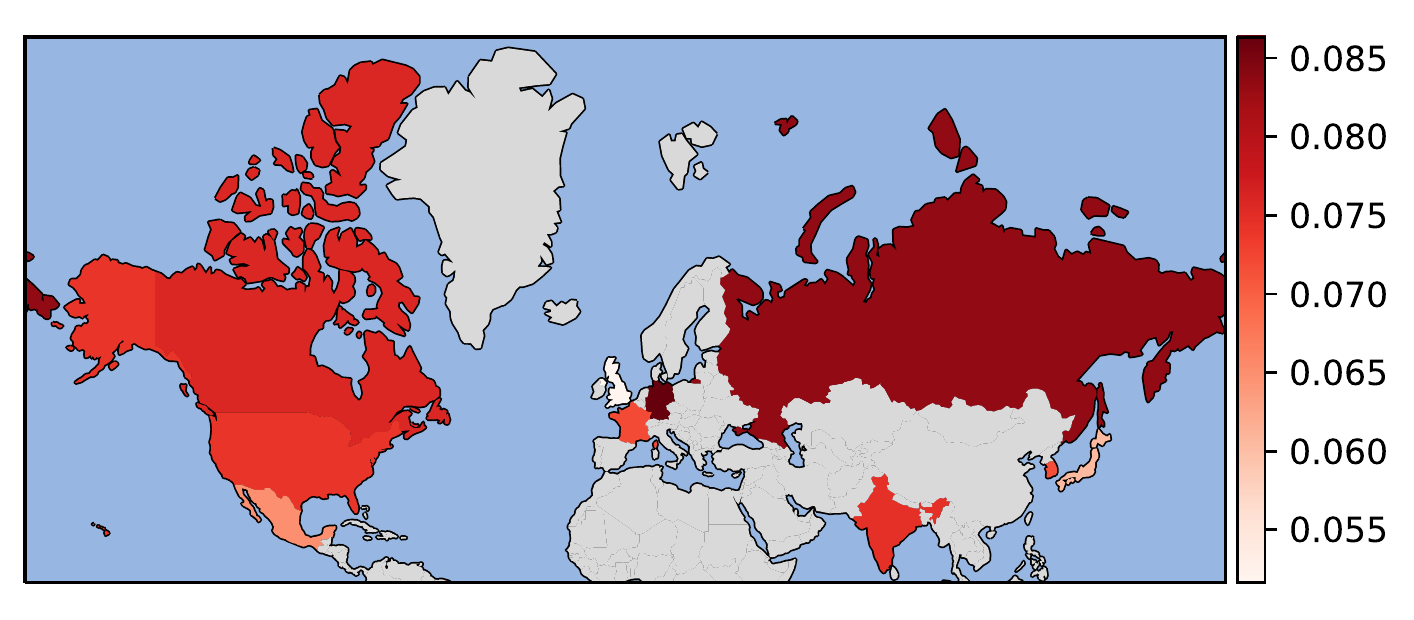} \\
(j) Snowboard & (k) Skateboard & (l) Surfboard \\
\includegraphics[width=5.6cm]{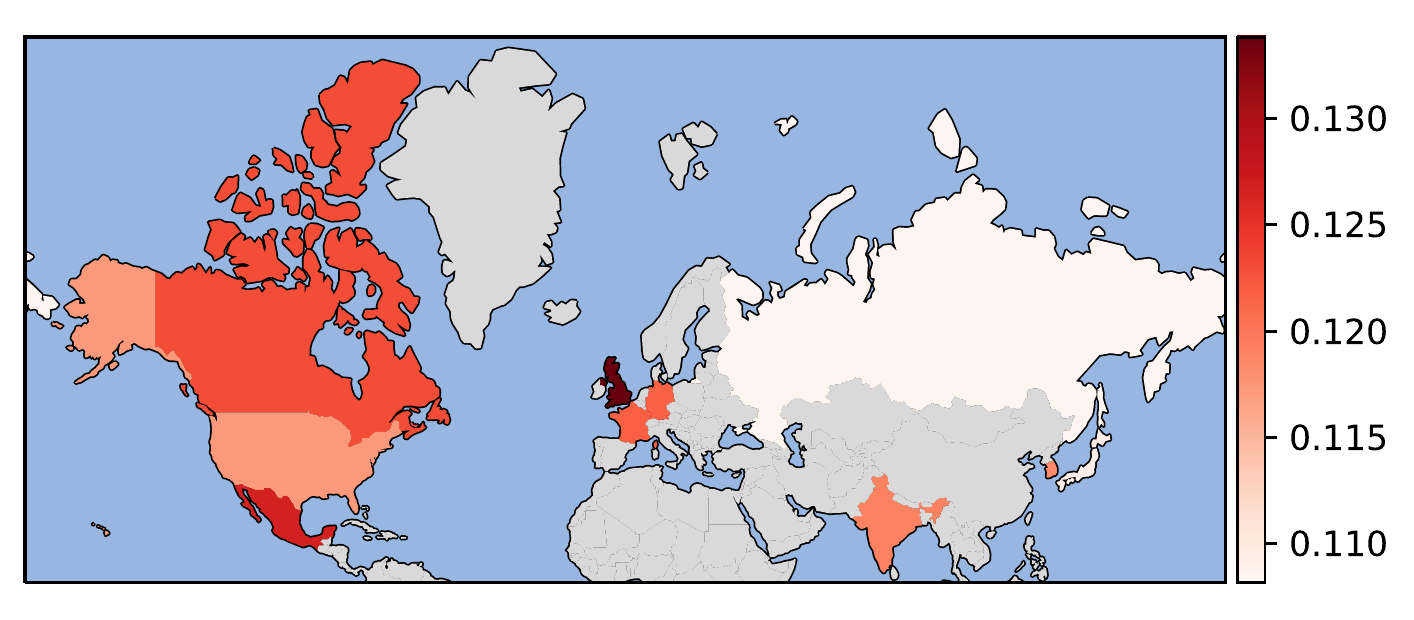} &
\includegraphics[width=5.6cm]{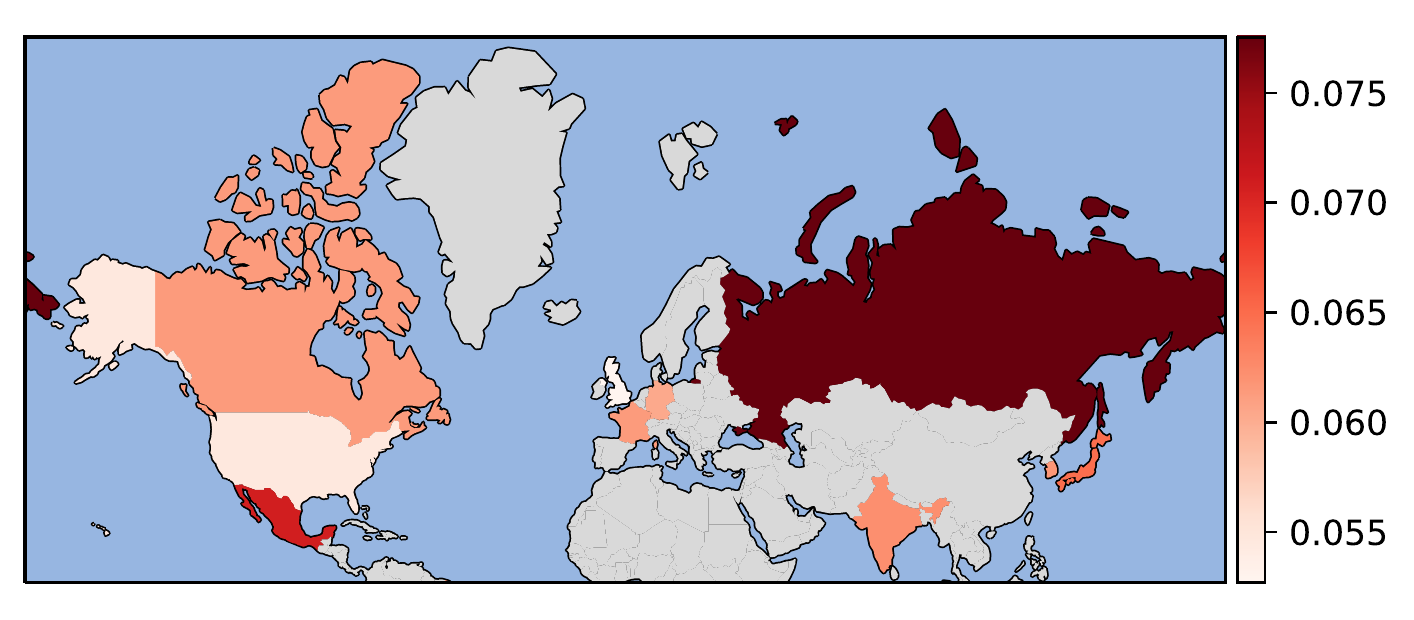} &
\includegraphics[width=5.6cm]{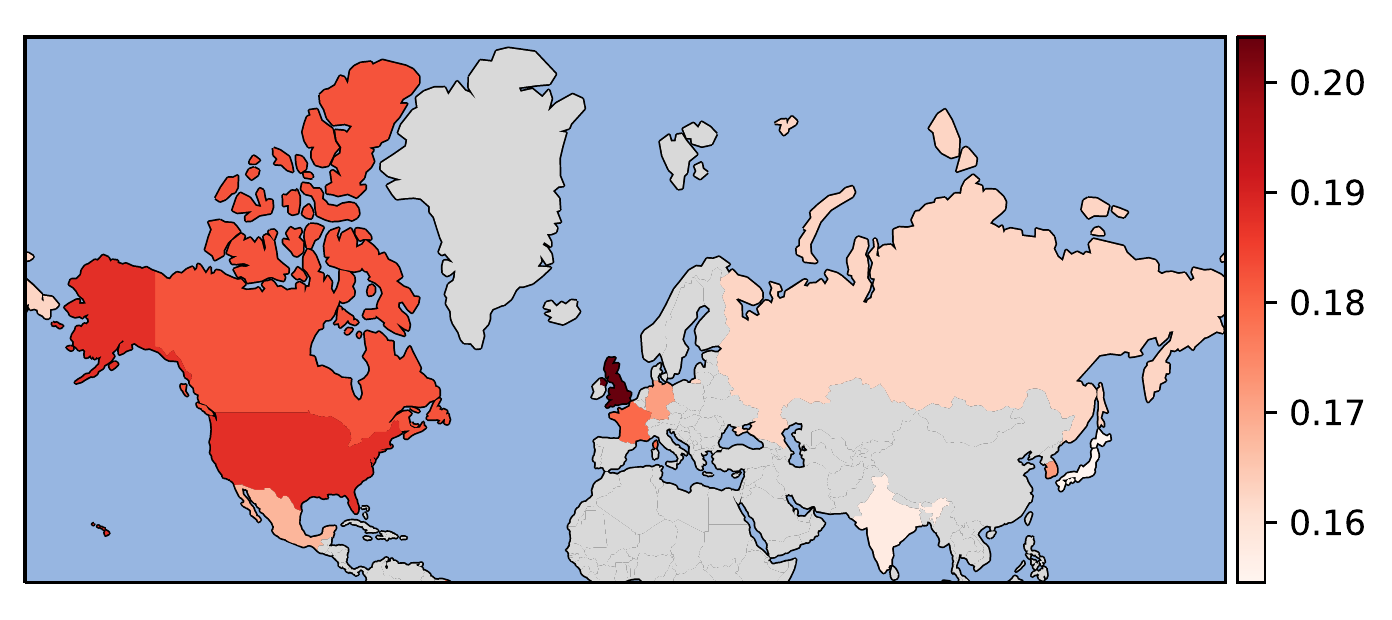} \\
(m) Tennis racket & (n) Kite & (o) Doughnut \\
\includegraphics[width=5.6cm]{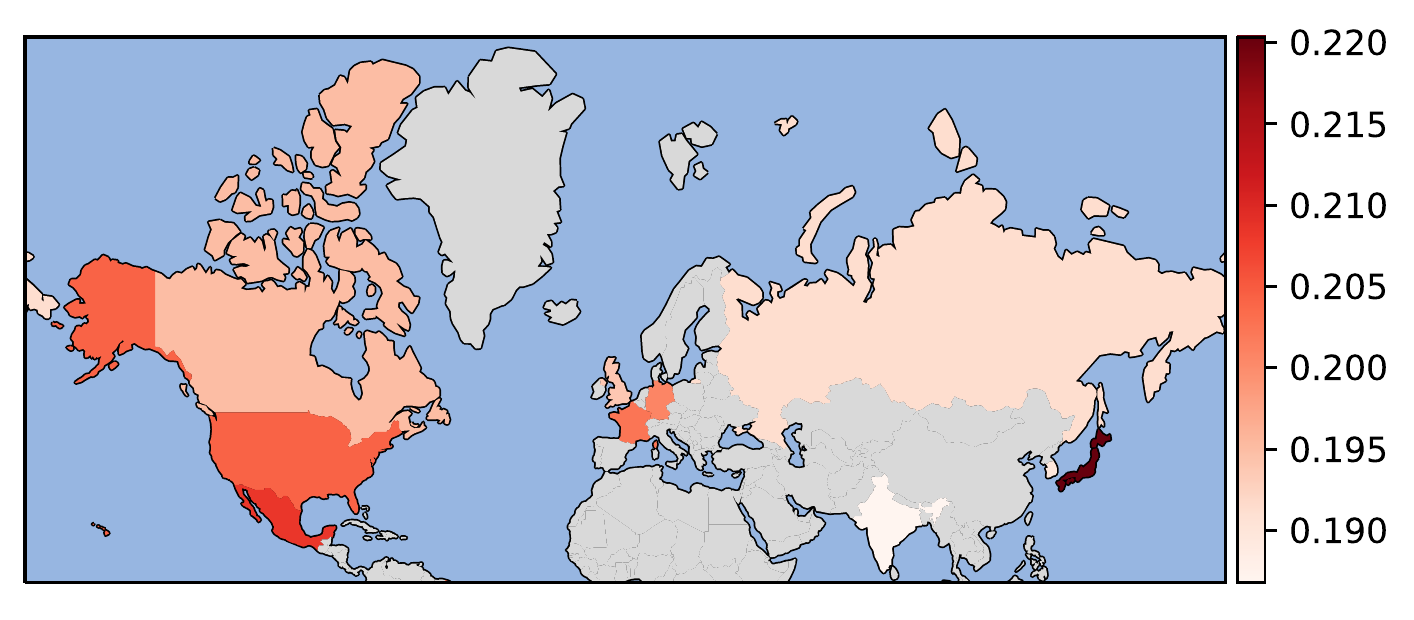} &
\includegraphics[width=5.6cm]{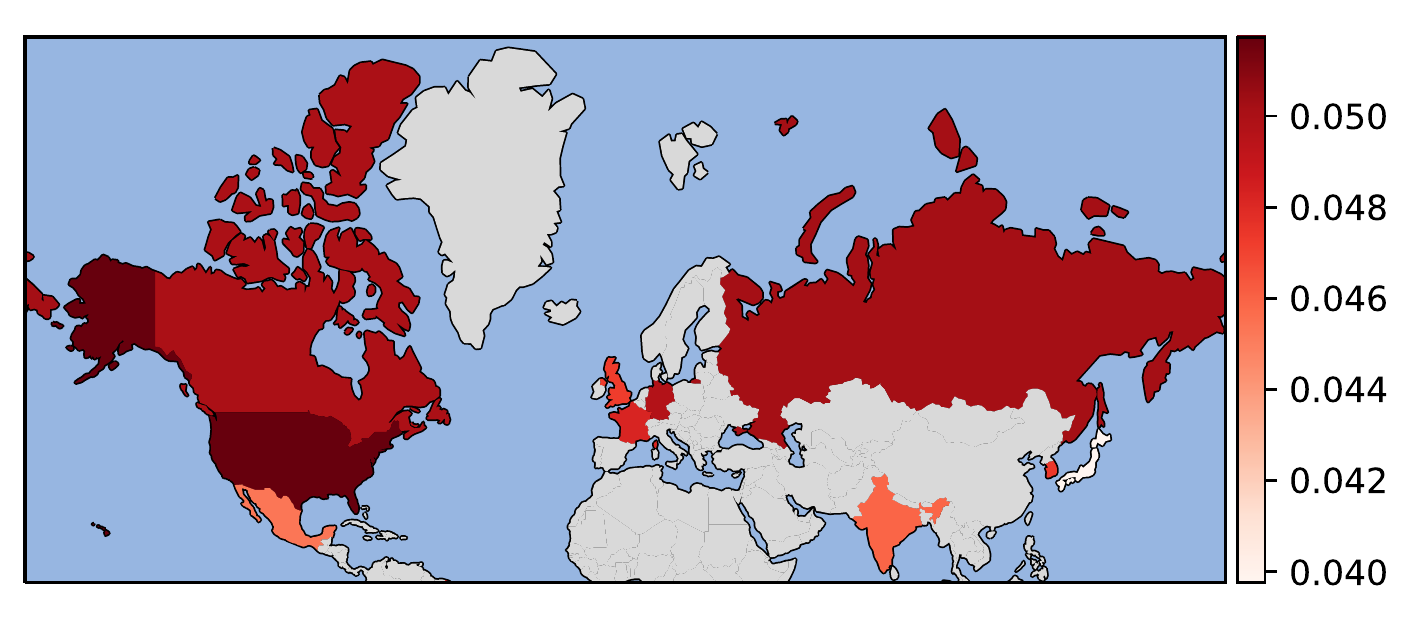} &

\includegraphics[width=5.6cm]{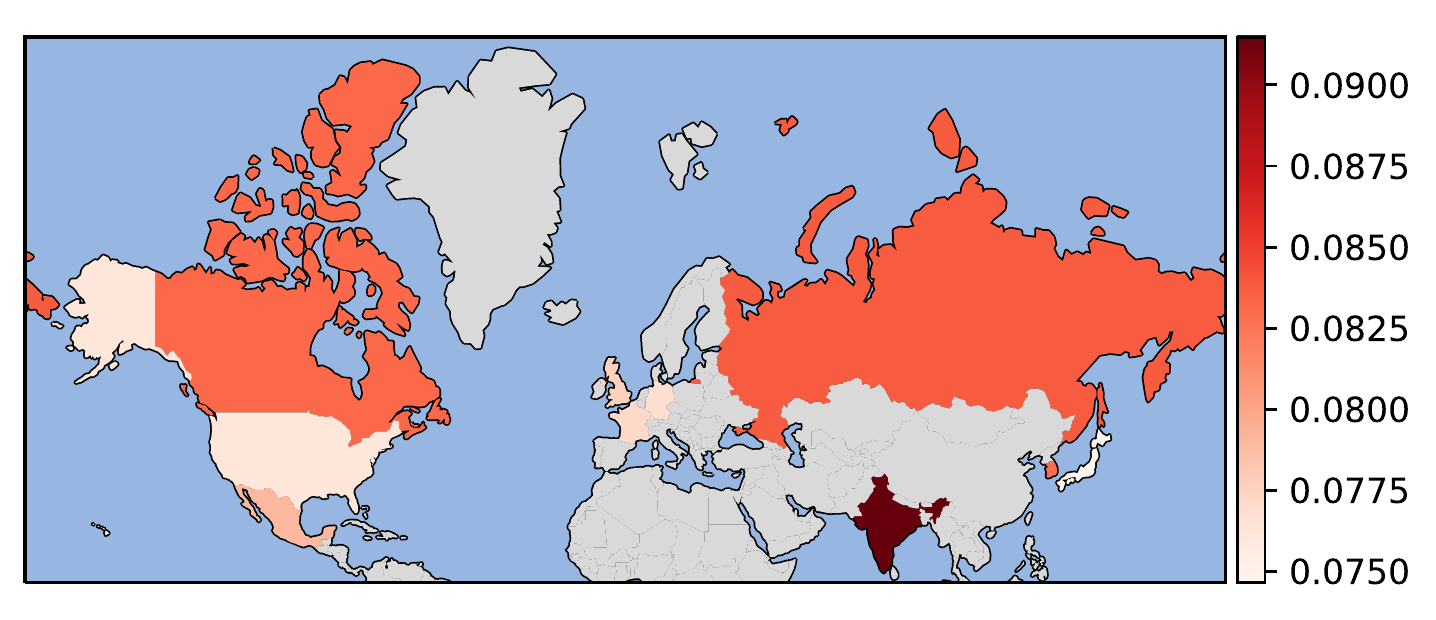} \\
(p) Cake & (q) Hot Dog & (r) Carrot \\
\includegraphics[width=5.6cm]{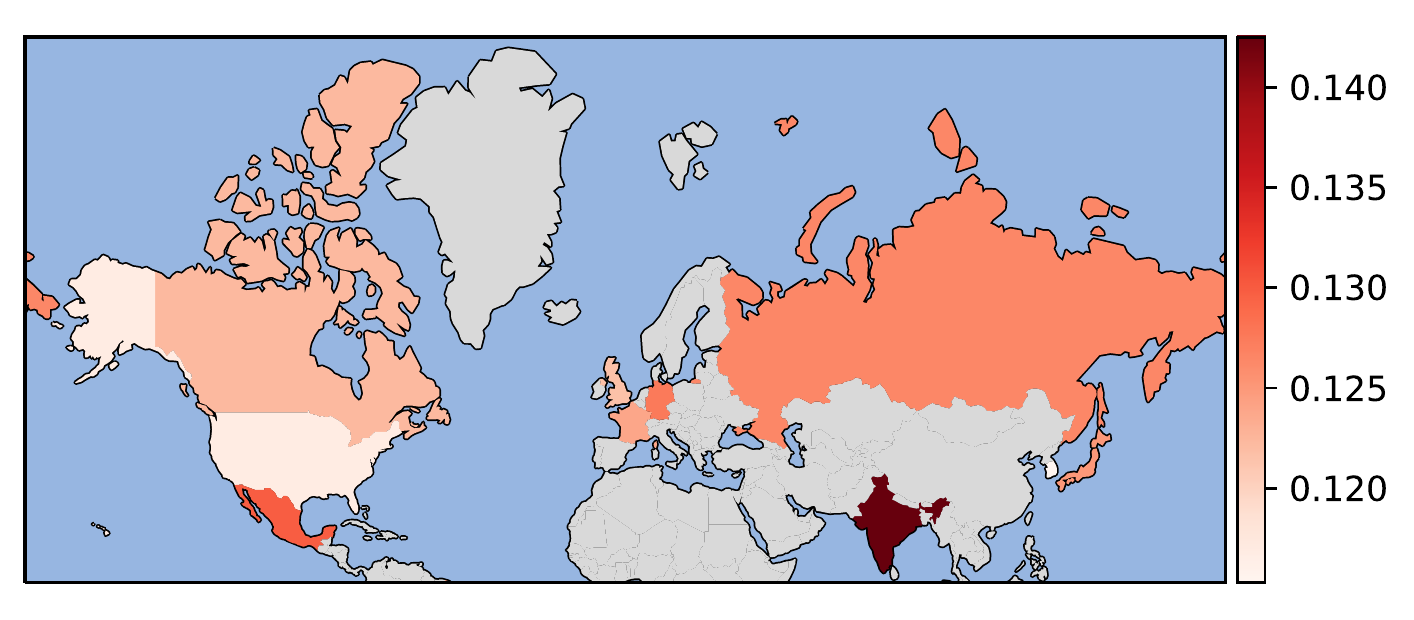} &
\includegraphics[width=5.6cm]{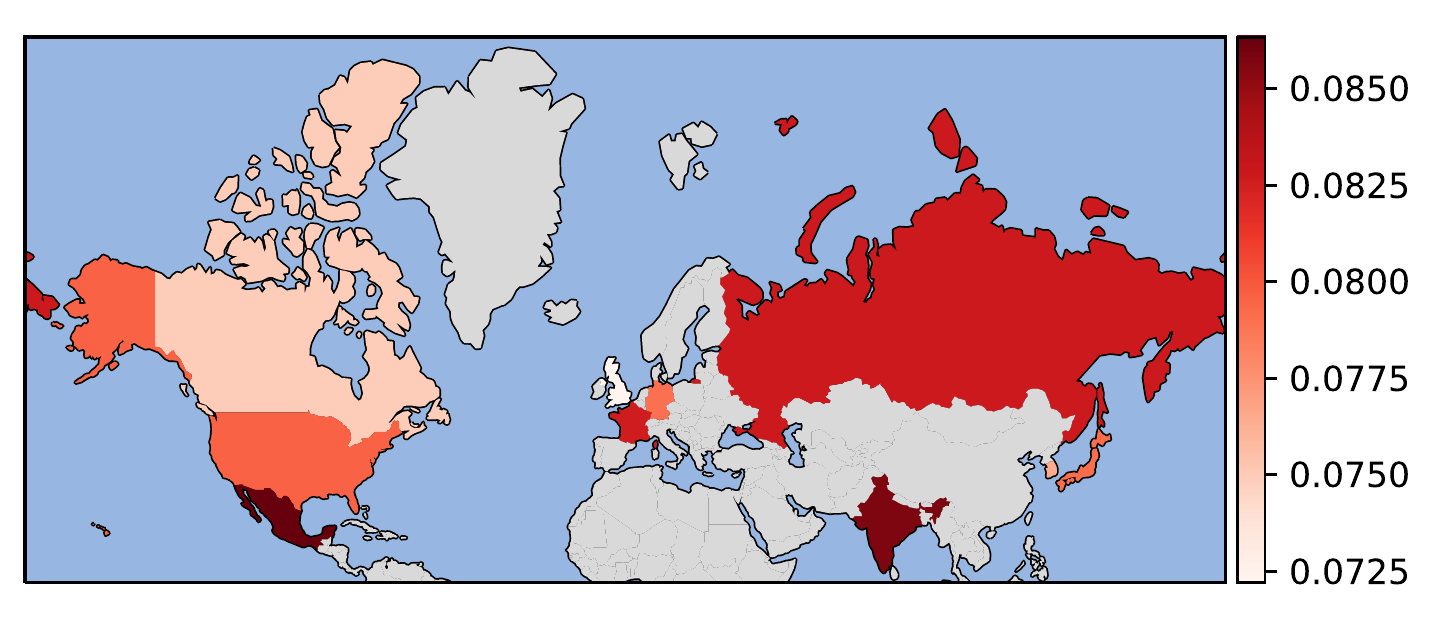} &
\includegraphics[width=5.6cm]{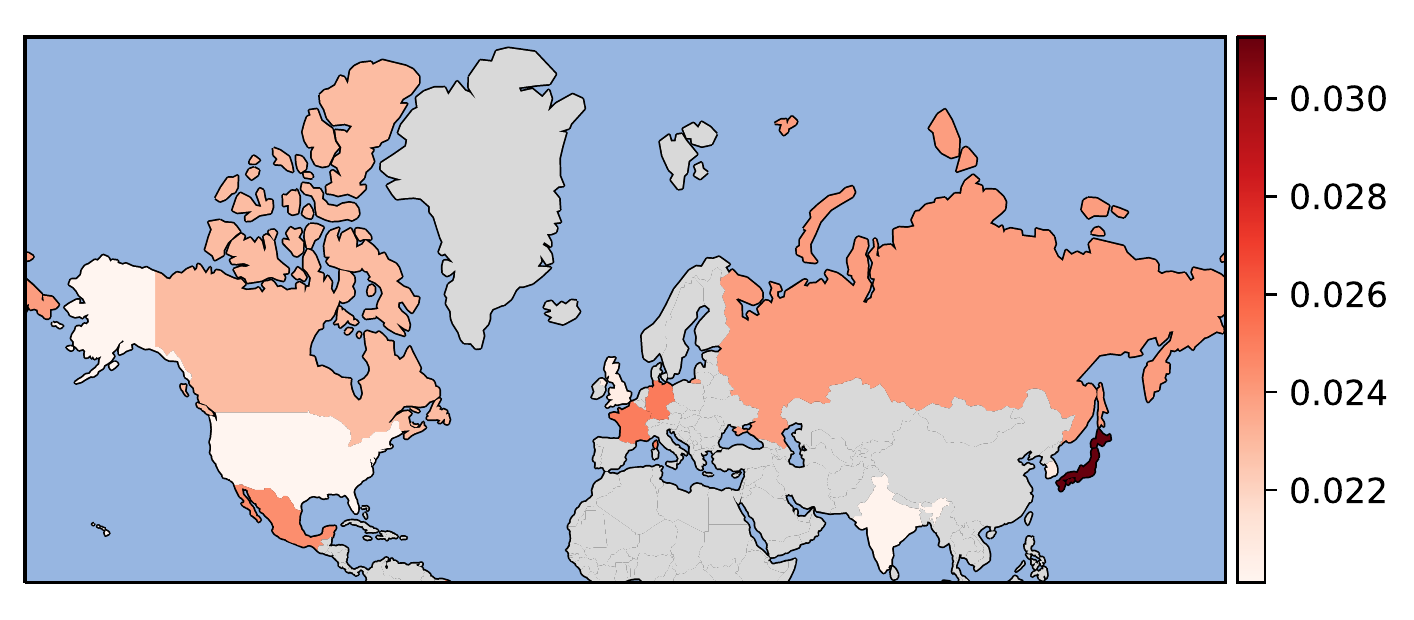} \\
(s) Broccoli & (t) Apple & (u) Banana\\
\end{tabular}
\caption{
Object distribution in the vehicle, sports and food categories across different countries. 
Red color represents the probability of an object occurring within a thumbnail.
The grey area are the countries where data are not collected.
Note that since both \textit{baseball bat} and \textit{baseball glove} represent the baseball sport, the total counts of these two labels is used for visualization.
}
\label{fig:geomap}
\end{figure*}

As shown in Figure~\ref{fig:examples}, the thumbnail images of the videos can provide rich object information regarding the videos, which may vary across the countries.
Indeed, this is the case since most trending videos bear different cultural preferences of different countries and such differences are reflected in the thumbnail contents.

In this section, we investigate three questions related to the objects in thumbnails. 
For each question, we also list several interesting observations from our experiment results.

\subsection{What is the distribution of the objects in the thumbnails in different countries?}

To answer this question, we first investigate the object distribution among all countries along with their geographical locations. 
We then measure the similarity between countries based on the object distribution.
We also perform an additional analysis on the correlation with some social indexes.

\subsubsection{Geographical object distribution} 
As depicted in Figure~\ref{fig:geomap}, certain objects tend to appear more frequently than other detected objects in each video.
Our analysis involves the most trending videos, thus the tendency towards certain objects reveals cultural preferences of the countries since the videos with such visual content are more favored by the large audience to make the video trending. 
Alternatively, Figure~\ref{fig:geomap} also allows us to draw similarities between the cultural preferences of different countries. Certain objects appear to occur more frequently together in certain countries. 

Looking closer in Figure~\ref{fig:geomap}, we can make several interesting observations.
First, when we compare the labels in ``Vehicle \& Transportations'' category, we find that the users from India have a higher interest in \textit{train} (Figure~\ref{fig:geomap} (f)), while the users in the  \textit{US} and \textit{Canada} are more interested in \textit{car} (Figure~\ref{fig:geomap} (d)).
When we compare the labels under the ``Sports'' category, we find that \textit{snowboard} (Figure~\ref{fig:geomap} (j)) has a larger popularity than \textit{skies} (Figure~\ref{fig:geomap} (i)) in a global trend. Meanwhile, \textit{skateboard} (Figure~\ref{fig:geomap} (k)) is more attractive in India and Japan.
Comparing the labels in the ``Food'' category, vegetables and fruits draw more attention from Indian users, while the western countries are more interested in the high-calorie foods like doughnuts, cakes and hot dogs.
Therefore, we can conclude that the similar frequency of occurrence among the most trending videos can indicate that the corresponding economies/countries share similar cultural preferences and characteristics in their consumption of YouTube content.

\subsubsection{Measuring similarity of countries} 
\begin{figure}[t!]
\centering
\begin{tabular}{cc}
\includegraphics[width=4.0cm]{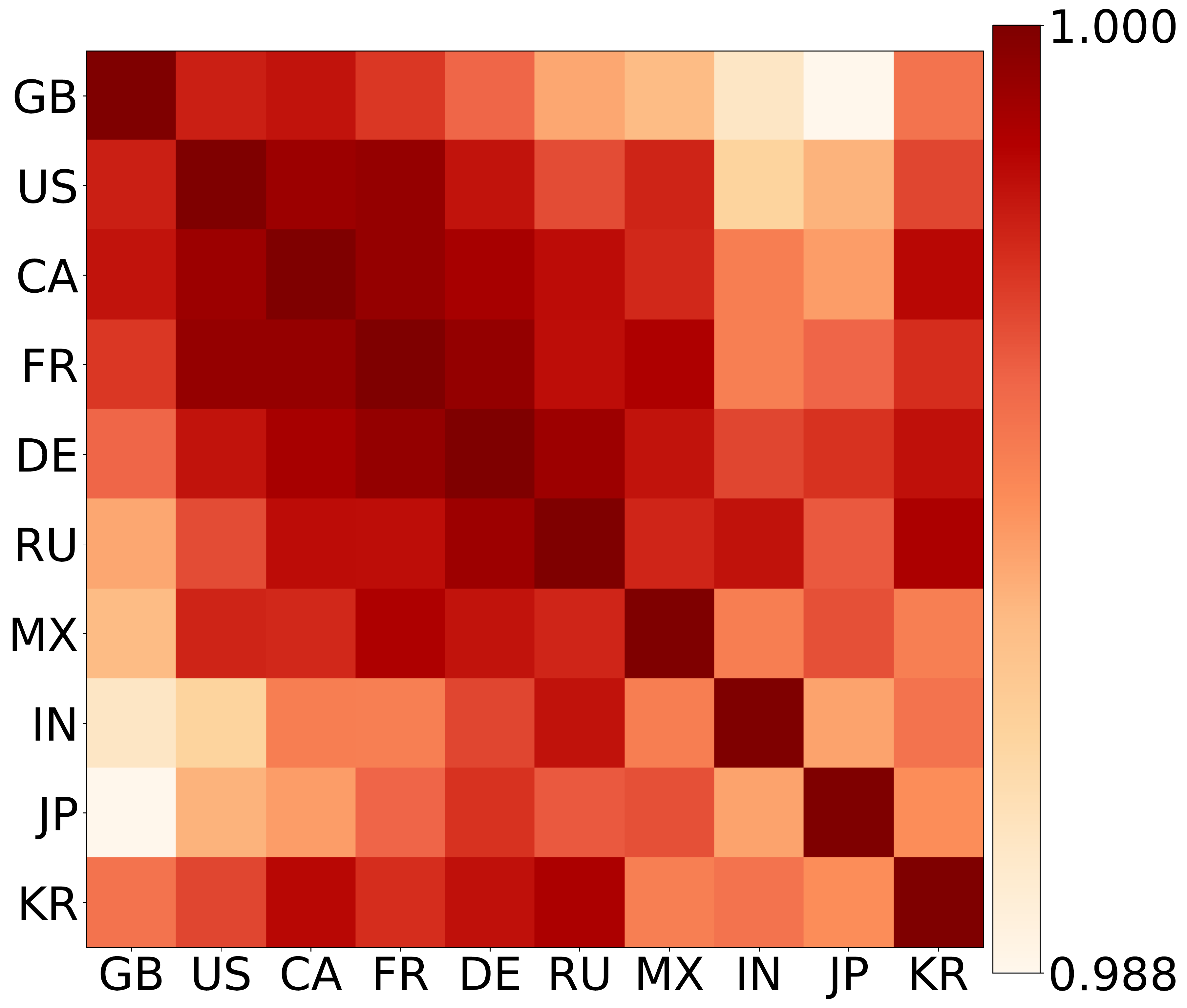} &
\includegraphics[width=4.0cm]{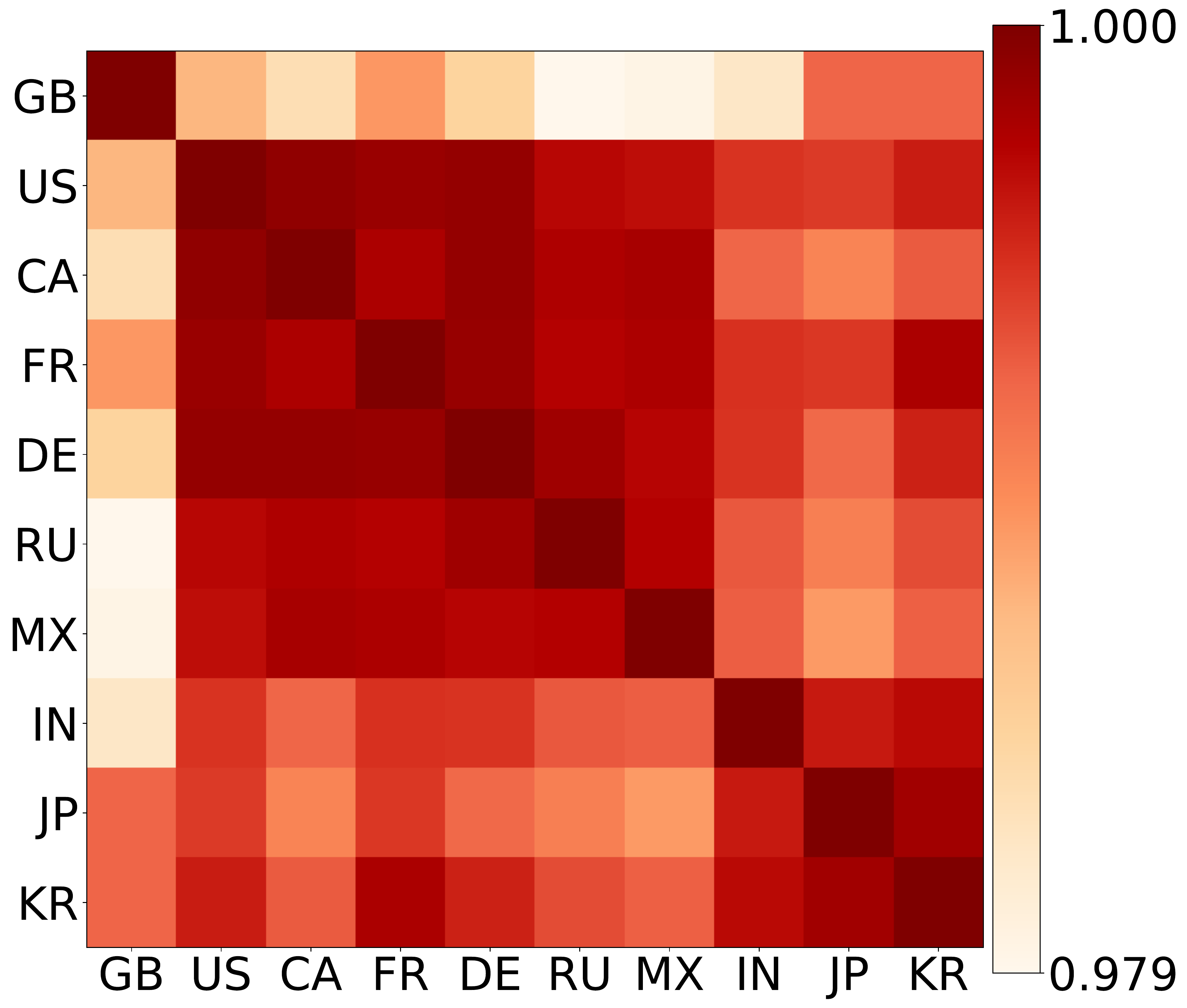} \\
(a) Transportation \& Traffic & (b) Sport \\ \includegraphics[width=4.0cm]{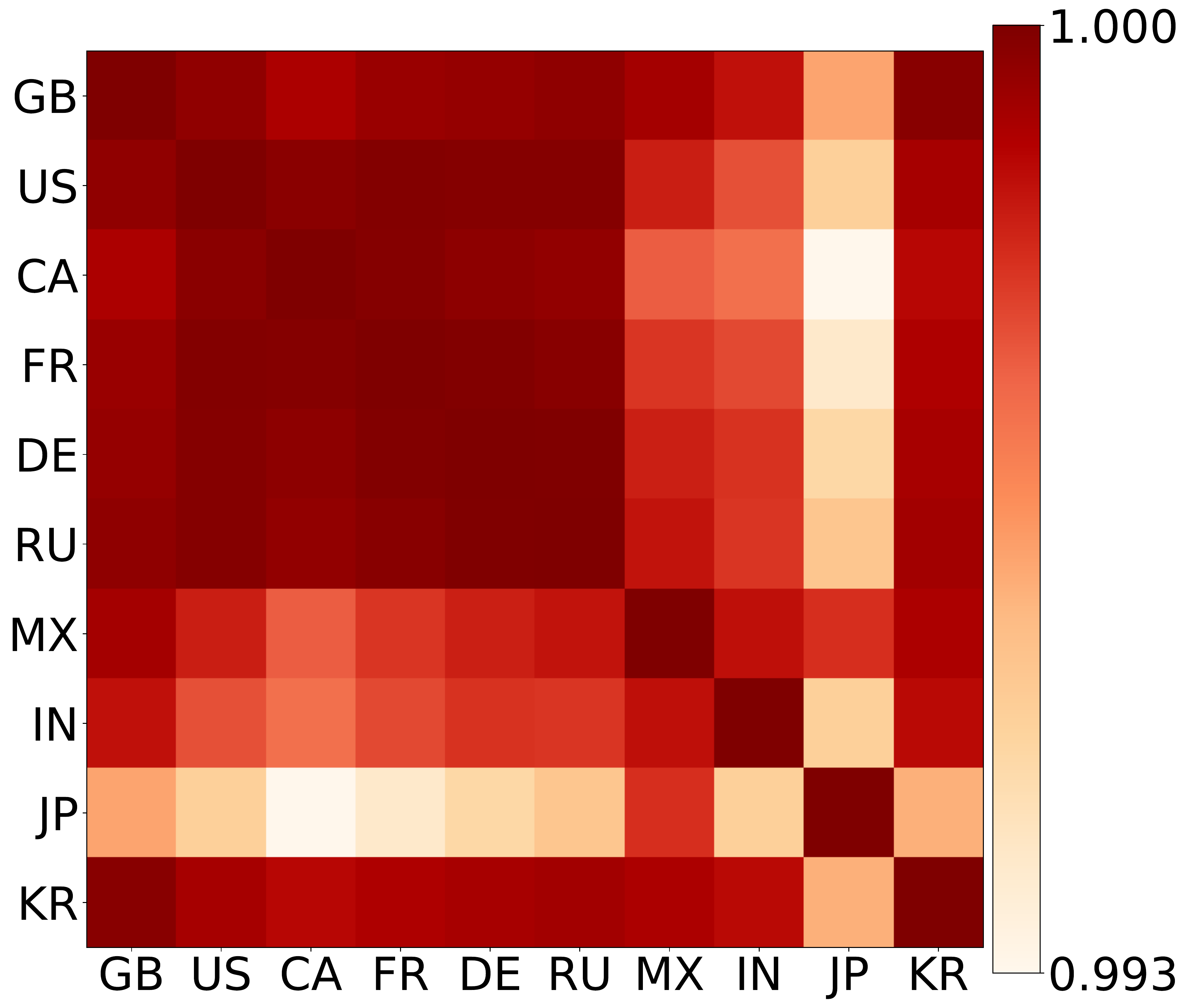} &
\includegraphics[width=4.0cm]{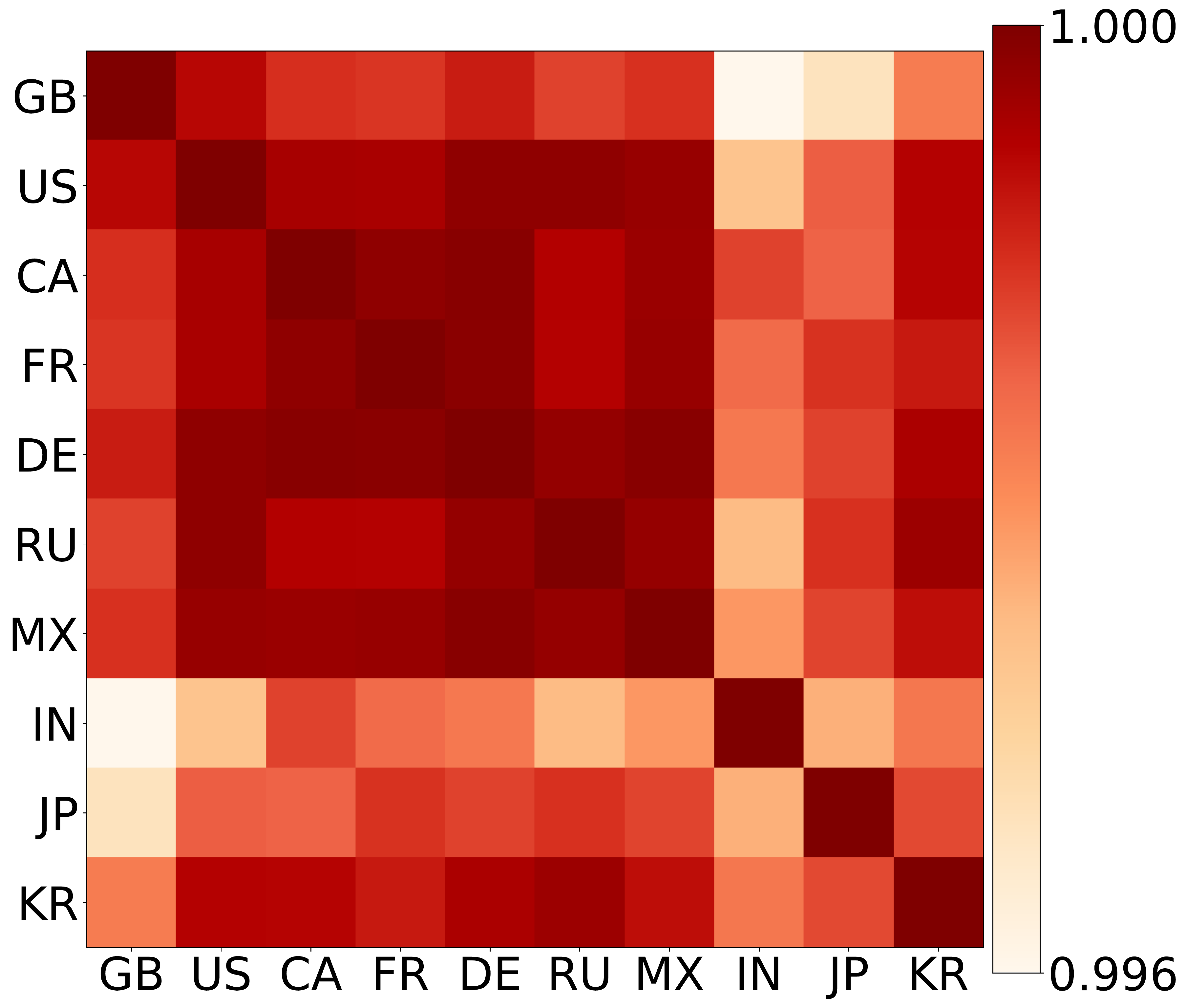} \\
(c) Food & (d) Animal\\
\end{tabular}
\caption{
Similarity of countries among different categories. Red color shade represents the extent of similarity between two countries.}
\label{fig:similarity}
\end{figure}

To gain a more comprehensive understanding of such preferences among different countries, we compute a similarity matrix based on the object distribution in each country, as shown in Figure~\ref{fig:similarity}.
The plots in the figure indicate the cosine similarity between the countries for the ratios of objects in different categories.
We observe that the \{\textit{US, CA, FR, DE, RU, MX}\} show higher correlation within their group in four categories, while \{\textit{IN, JP, KR}\} only show high correlation in the ``Sports'' category (Figure~\ref{fig:similarity} (b)).
The correlational similarity may be attributed to the popular sports preferences of the audiences, as well as the  geo-cultural proximity of the countries. 
Therefore, the ratios of objects tend to vary together for the groups of countries in different categories, signaling a cultural proximity in their preference of video content consumption.

\begin{table}[t!]
    \centering
    \begin{tabular}{|c|c|c|}
    \hline
    Social Index & $r$ & $p$-value \\
    \hline    
    Population & $-0.405$ & $2.94\times 10^{-5}$ \\
    Life Expectancy & $-0.448$ & $3.00\times 10^{-6}$ \\
    HDI & $-0.434$ & $6.48\times 10^{-6}$ \\
    GDP per Capita & $-0.366$ & $1.79\times 10^{-4}$ \\
    Literacy Rate & $-0.421$ & $1.28\times 10^{-5}$ \\
    \hline
    \end{tabular}
    \caption{Correlations between our similarity matrix and social-economic statues between countries.}
    \label{tab:social_index}
\end{table}

\subsubsection{Measuring correlation with social indexes} Following previous work~\cite{you2017cultural,park2017cultural}, we also examine whether our visual similarity is correlated with some widely recognized social indexes\footnote{Data are collected from http://www.aneki.com/comparison.php}, such as population, life expectancy, HDI (i.e. human development index), GDP per Capita, and literacy rate.
In more details, we use the absolute difference between the social indices of a pair of countries to measure their socio-economic closeness.
Therefore, if the social index difference between the countries is small, then these two countries are more similar socially or economically.
Next, we utilize the Pearson correlation coefficient to measure the correlation between our visual similarity matrix with the social index difference, as shown in Table~\ref{tab:social_index}.
Interestingly, we find that the visual similarity is correlated with all social indexes with a high confidence ($p$-value $<2e-4$ for all social indexes), although none of the visual attributes yields a strong correlation with them.
This suggests that there are still underlying cultural factors within videos that have not been discovered yet, or the linear correlation measurement is not ideally suitable for this problem. 

\begin{figure}[!tb]
    \centering
    \includegraphics[width=0.5\textwidth]{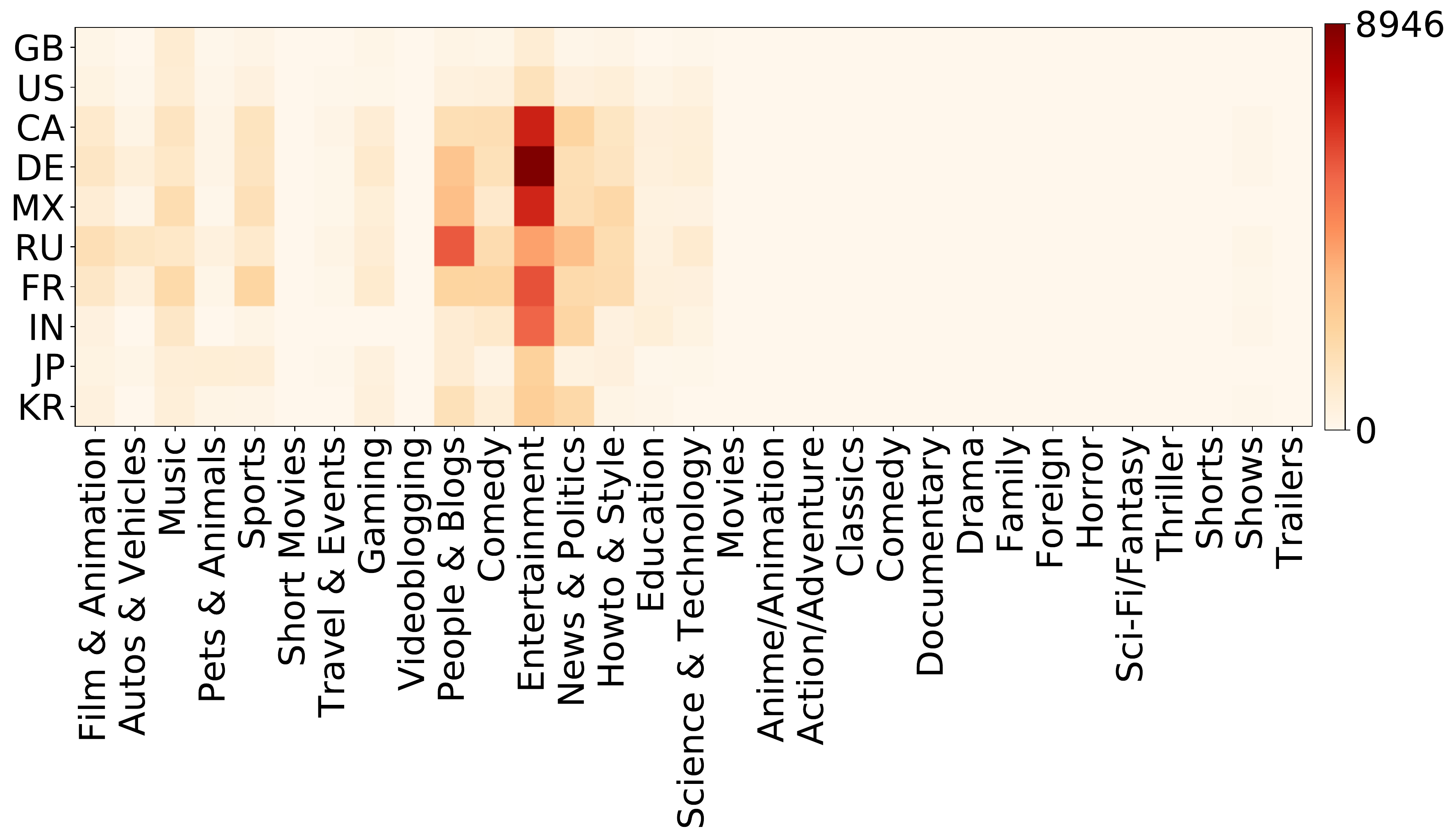}
    \caption{The video count based on countries and genres.}
    \label{fig:country-genre}
\end{figure}

\begin{figure*}[!tb]
    \centering
    \includegraphics[width=\textwidth]{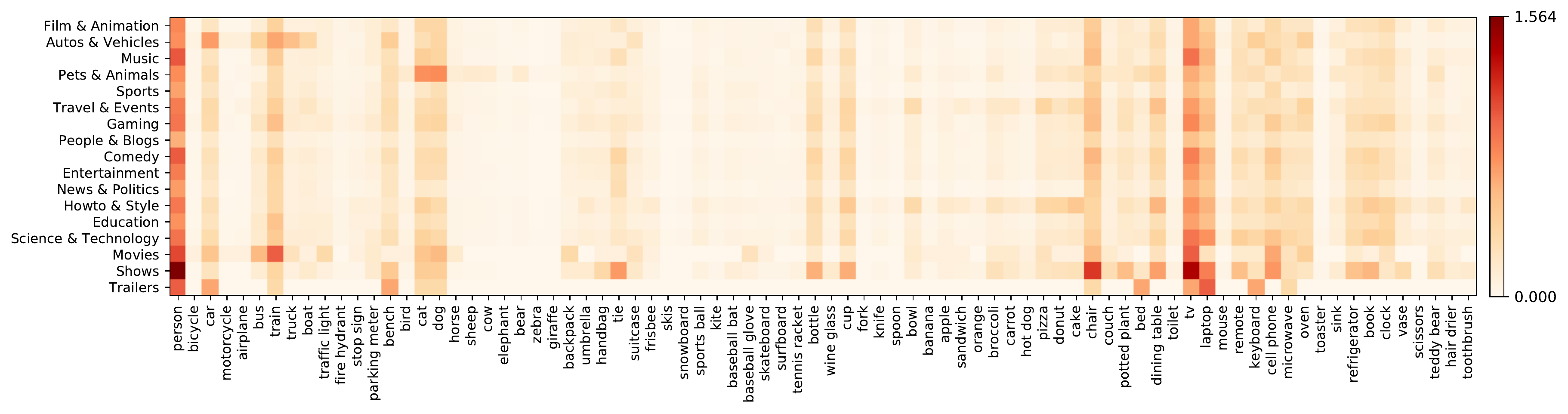}
    \caption{The object distribution for different genres in the US. Several genres are ignored in this figure due to their low counts. In each row, the object counts are divided by the total number of videos in each genre.}
    \label{fig:genre-object}
\end{figure*}

\begin{figure*}[t!]
\centering
\begin{tabular}{cccc}
\includegraphics[width=4.0cm]{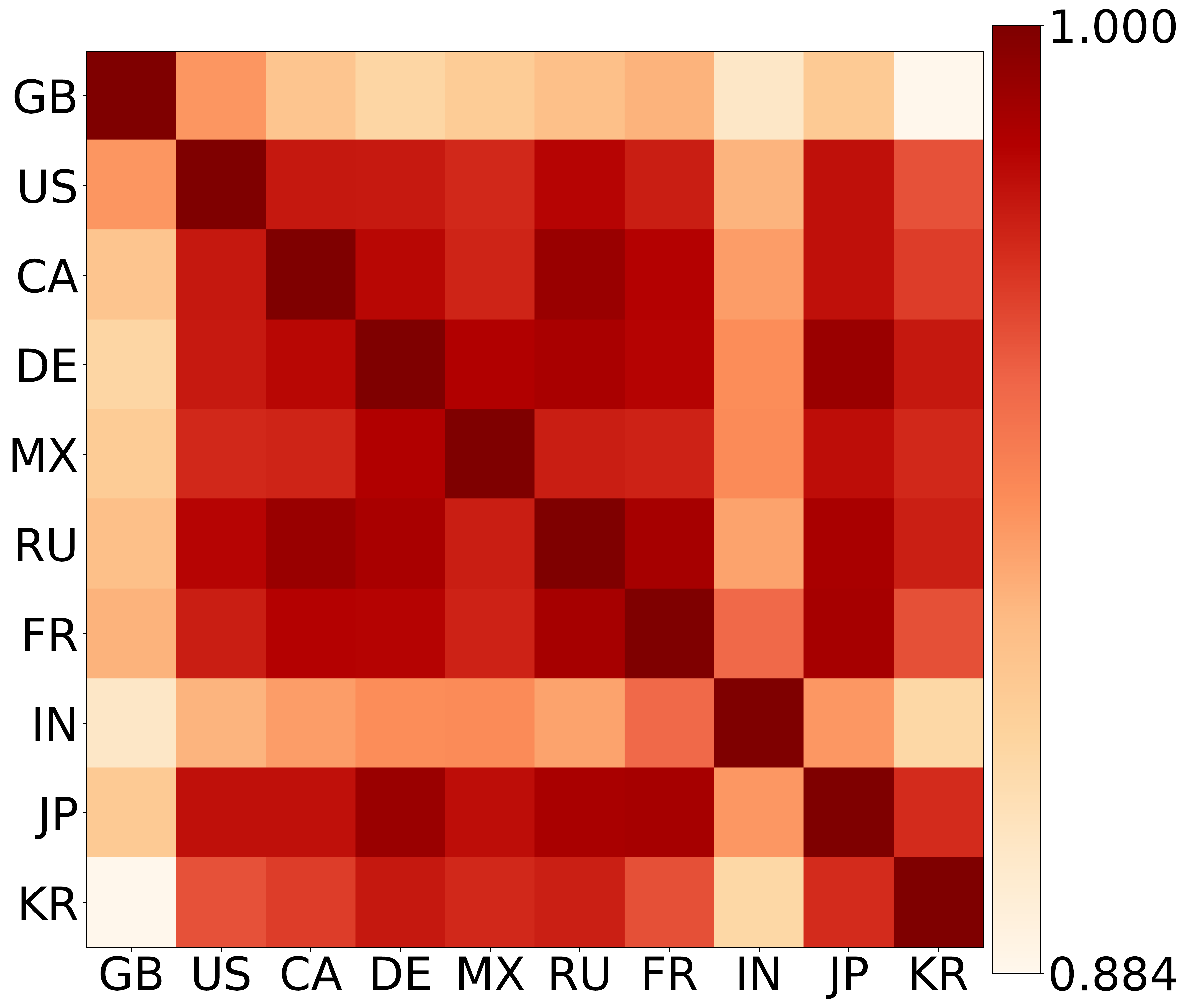} &
\includegraphics[width=4.0cm]{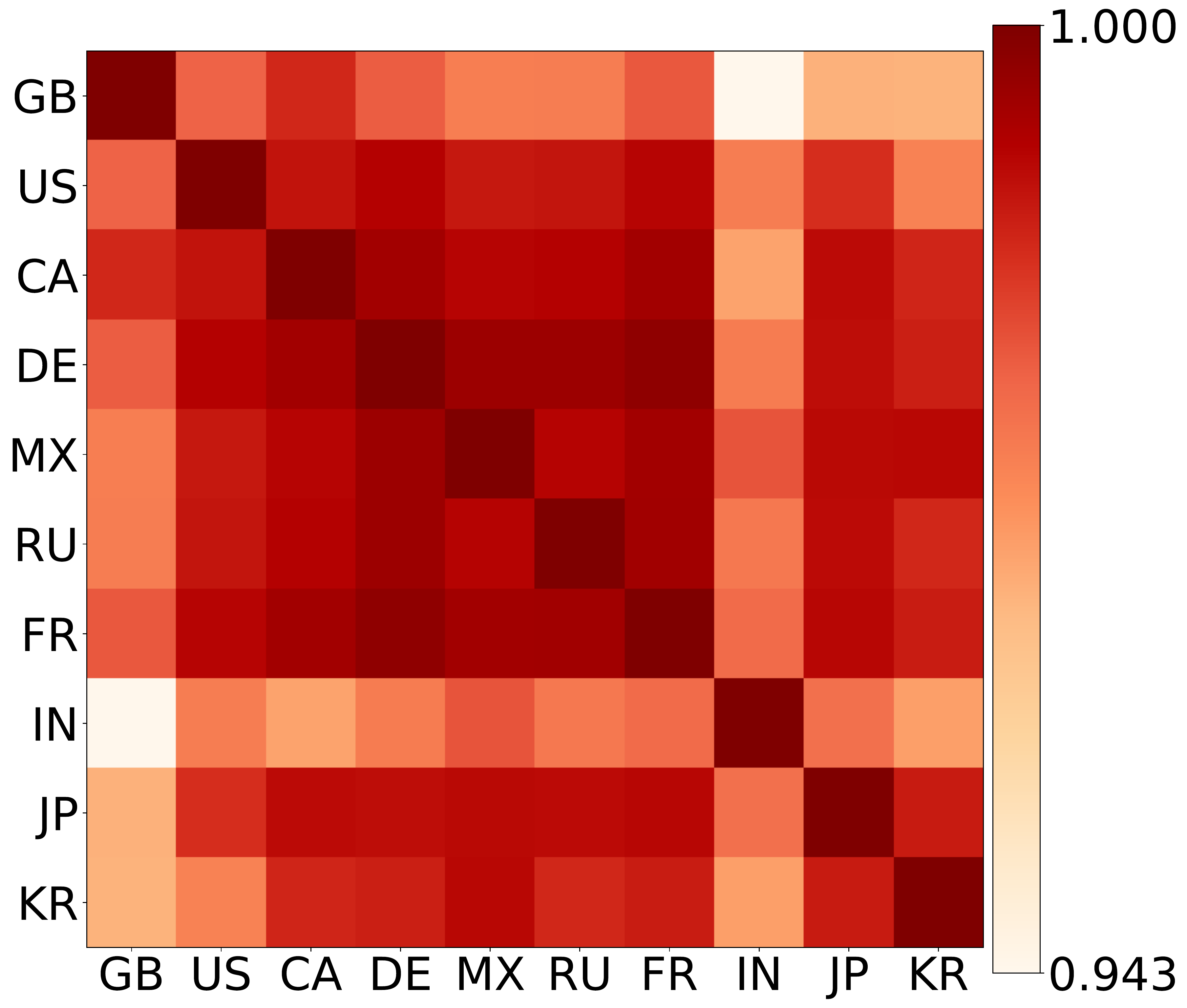} &
\includegraphics[width=4.0cm]{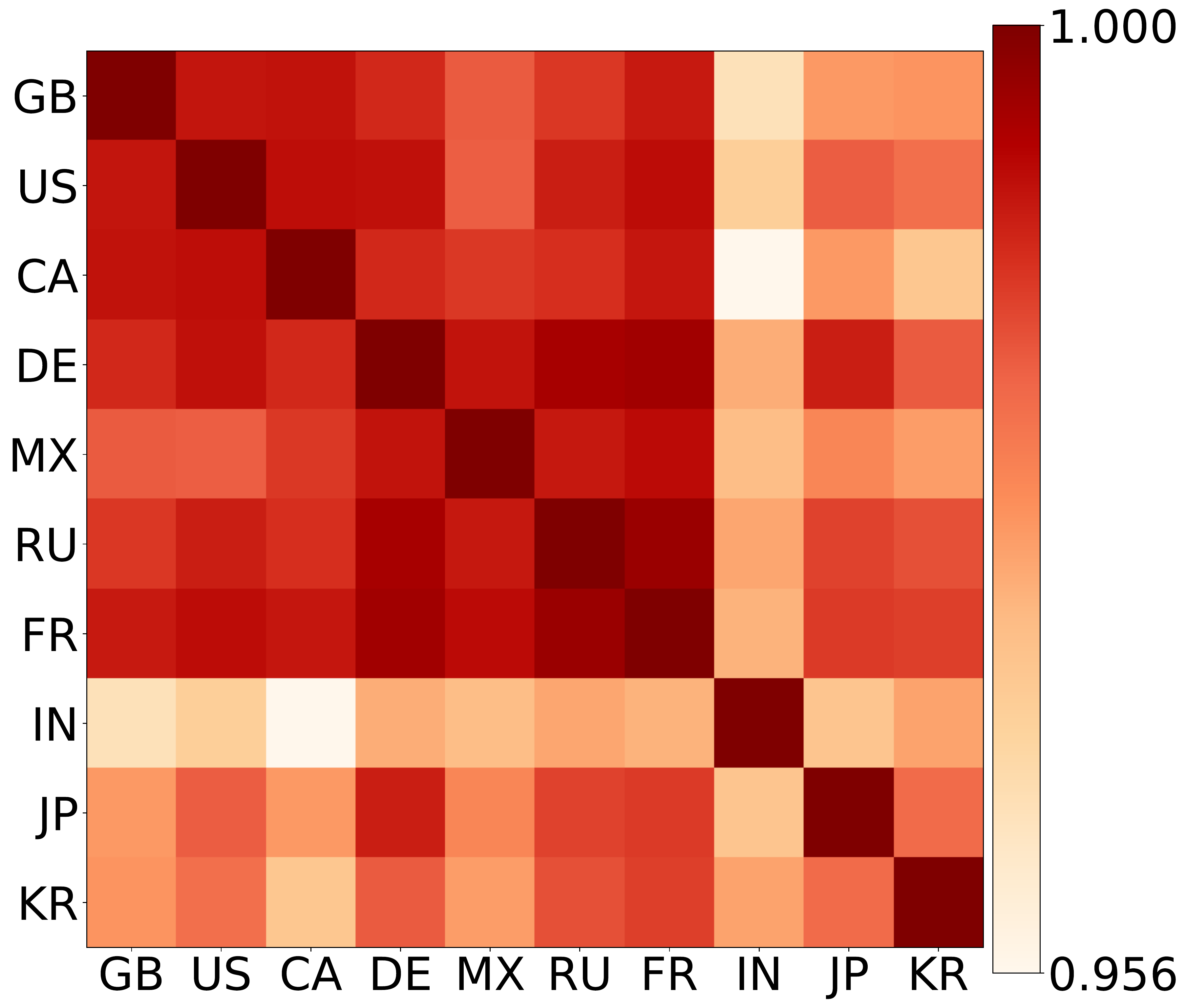} &
\includegraphics[width=4.0cm]{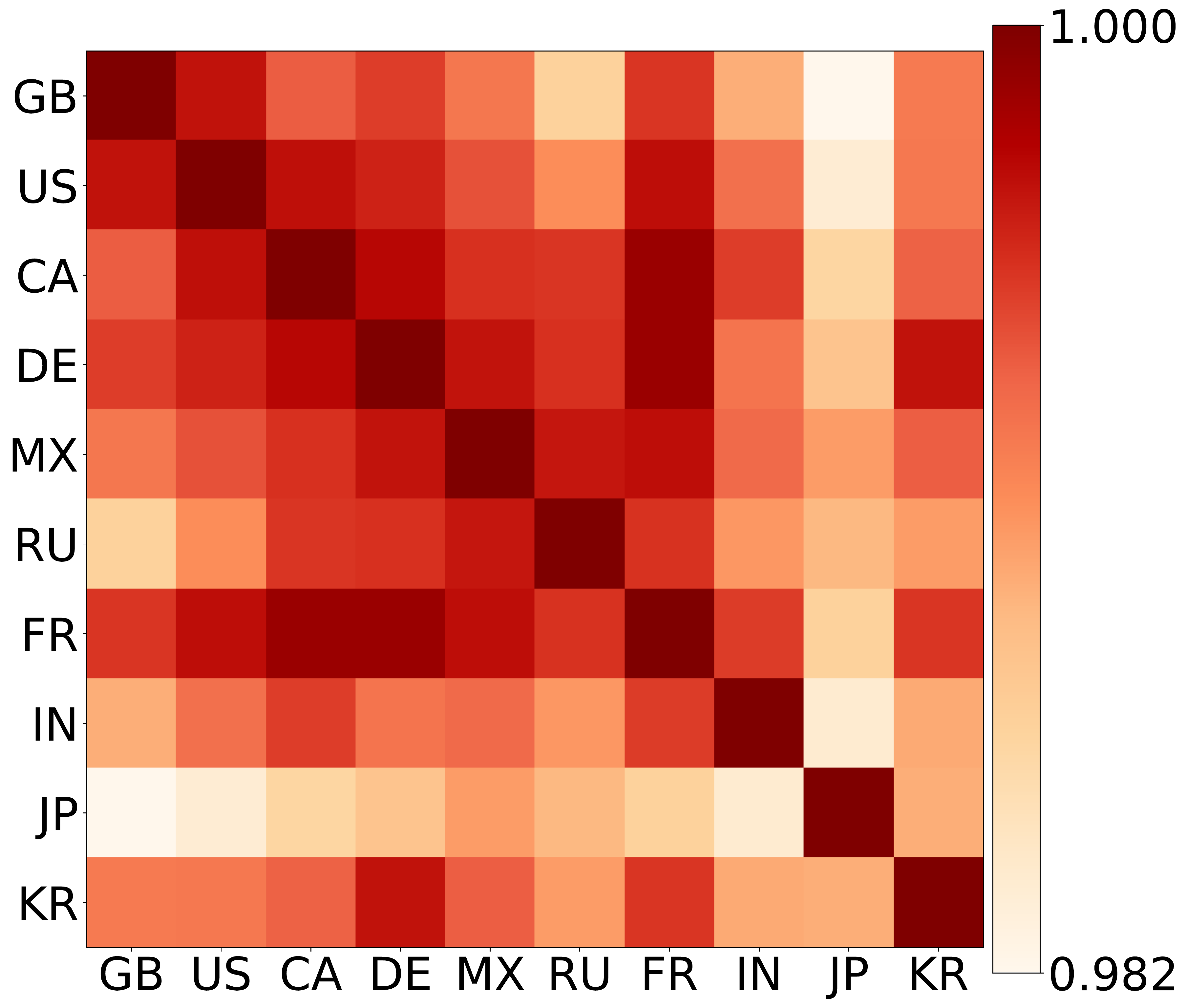} \\
(a) Autos \& Vehicles & (b) News \& Politics & (c) Sports & (d) Entertainment\\
\includegraphics[width=4.0cm]{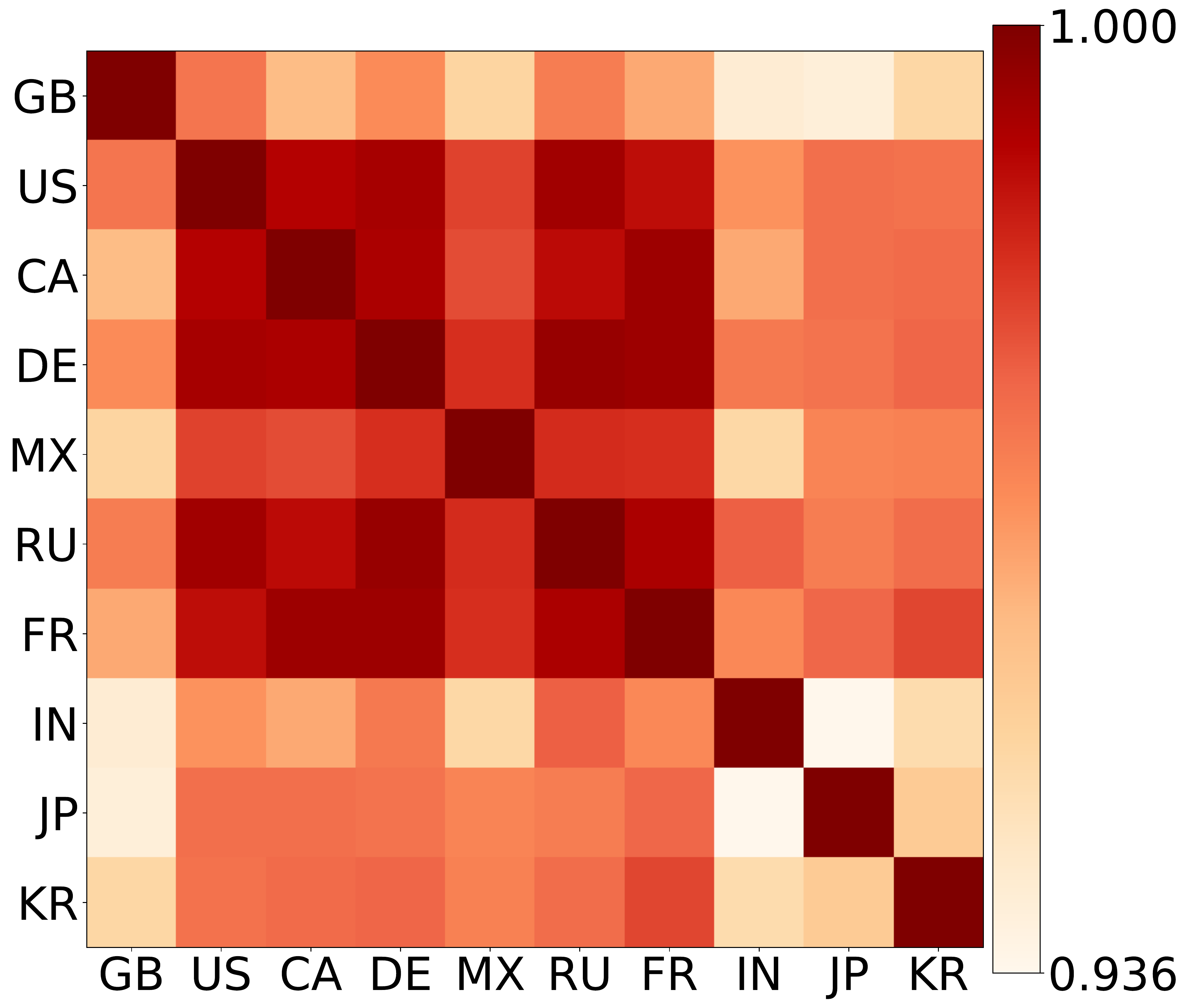} &
\includegraphics[width=4.0cm]{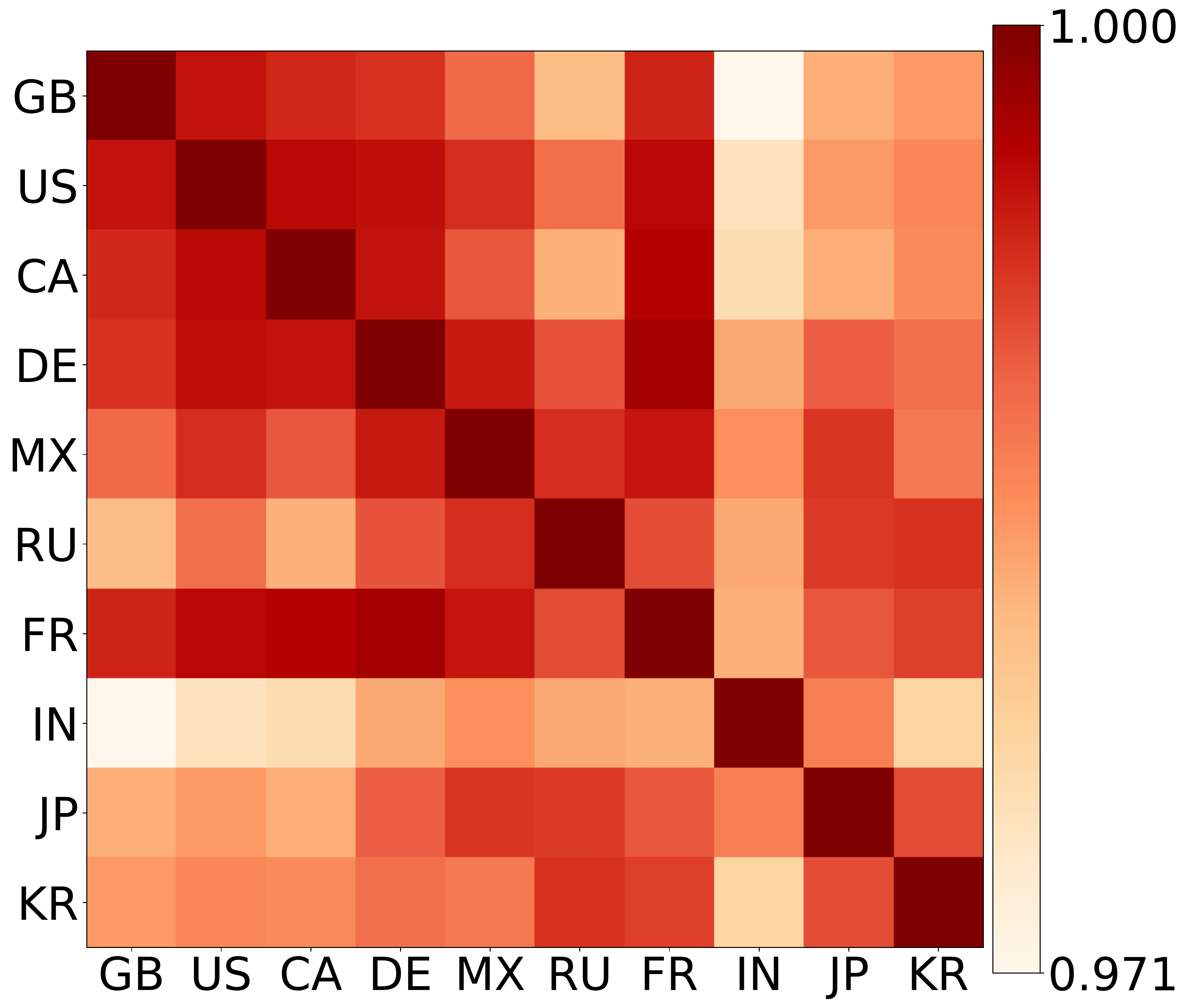} &
\includegraphics[width=4.0cm]{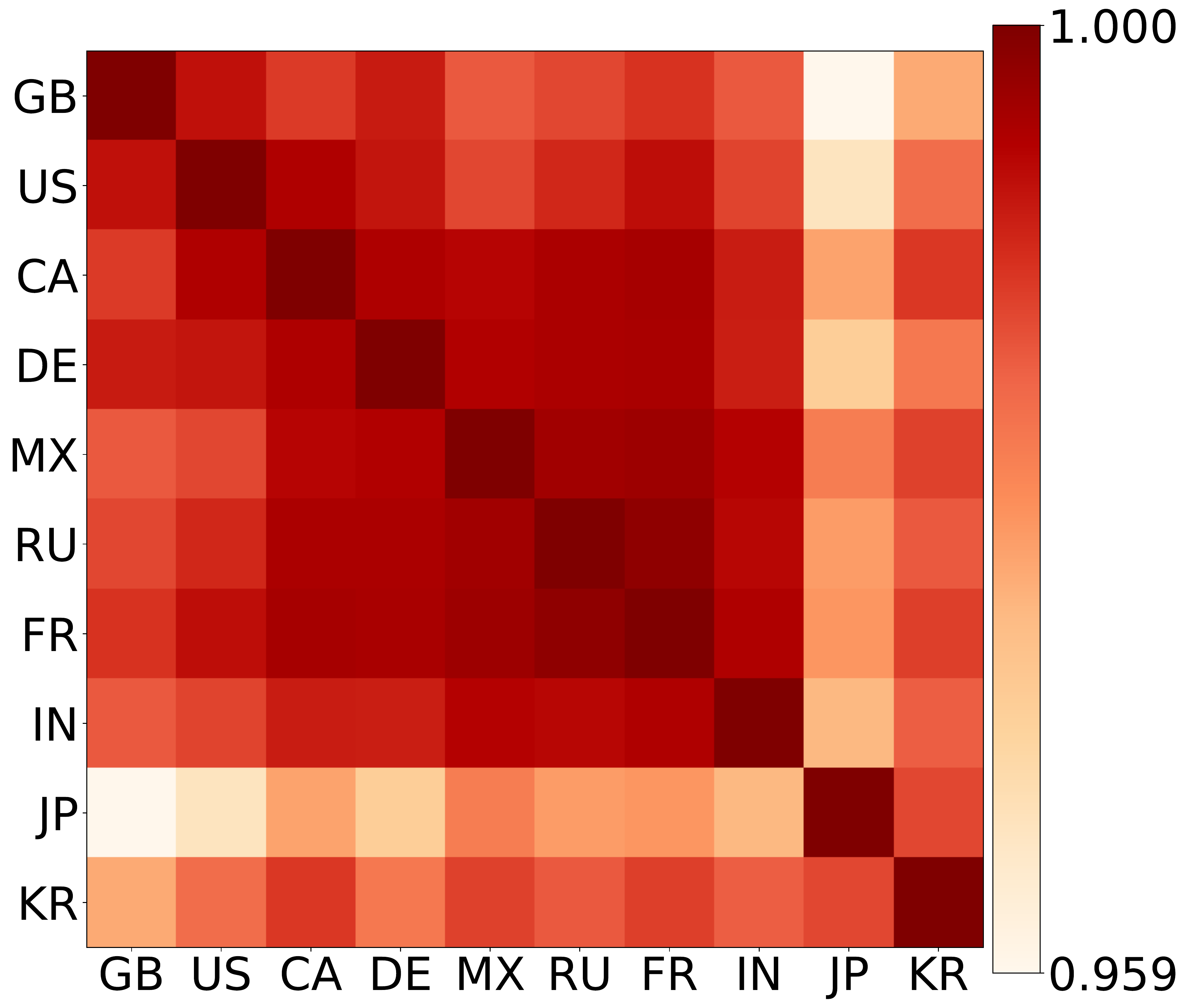} &
\includegraphics[width=4.0cm]{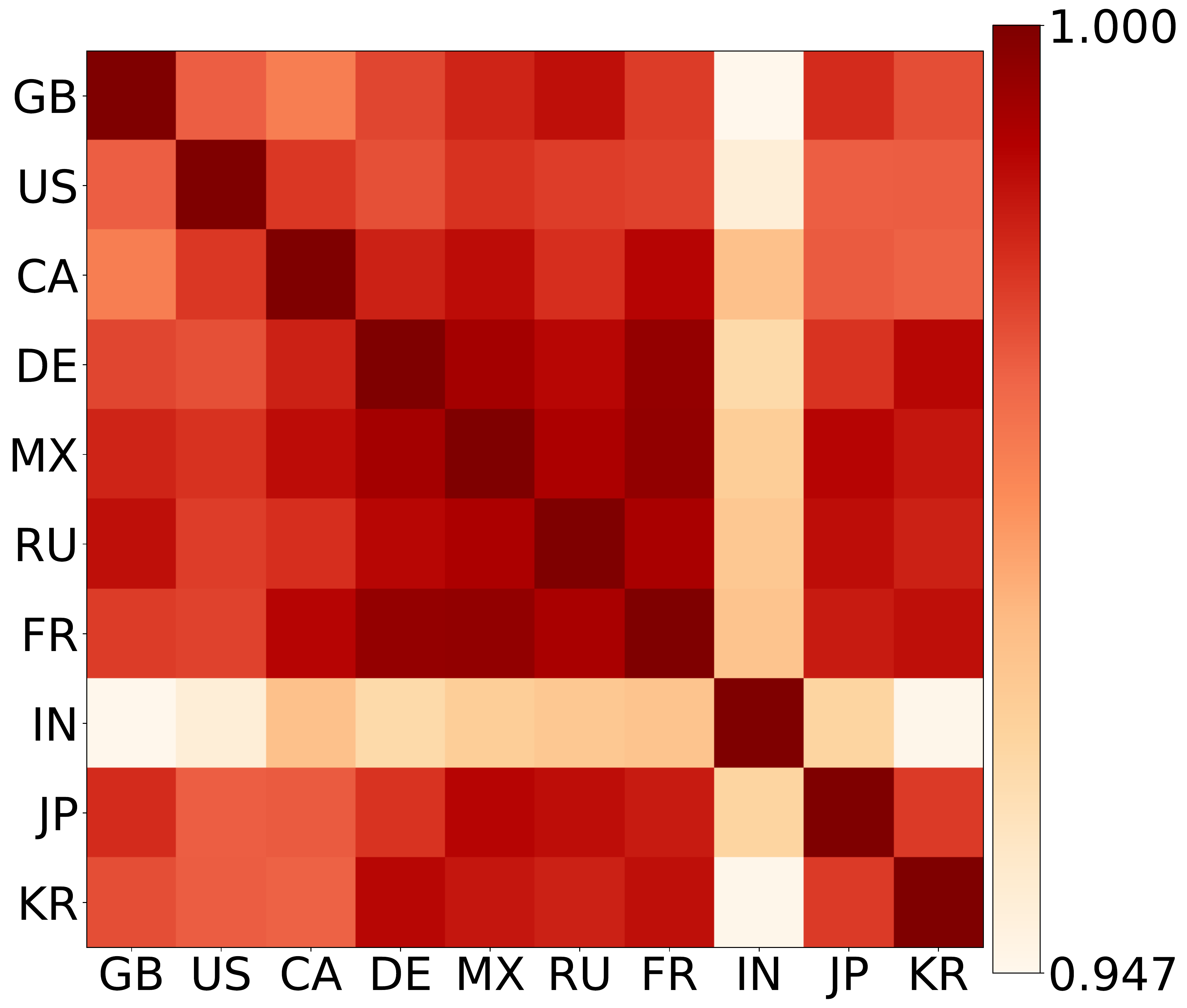} \\
(e) Science \& Pets \& Animals & (g) Comedy & (h) Gaming\\
\end{tabular}
\caption{
Similarity of object-genre relationship among different countries.}
\label{fig:obj_channel_country_similarity}
\end{figure*}

\subsection{Is the correlation between objects in thumbnails  and video genres same among different countries?}

To answer this question, we first explore the popularity of the genres in different countries. 
Then we investigate the correlation between the video genres and the objects appearing in the thumbnails. 
Finally, we demonstrate the similarity of such correlation between different countries.

\subsubsection{Measuring the popularity of video genres}

First, we explore the popularity of different video genres by counting the number of trending videos for each genre and country, as shown in Figure~\ref{fig:country-genre}.
In general, most countries like \textit{entertainment} videos the most. Especially \textit{GE} occupies the largest portion in entertainment videos compared to other countries.
The only exception in those ten countries is \textit{RU}, where ``People \& Blogs'' is the most popular video genre.
There are also some other interesting findings when we look at  specific genres. For example, the videos under the  ``Politics'' genre are less trending thus perhaps less cared for in \textit{JP} and \textit{GB} in contrast to \textit{CA} and \textit{RU}. The YouTube audiences in \textit{JP} and \textit{MX} favor watching the videos under ``Pets and Animals'' more than that in \textit{IN}.
We also observe that the movies, trailers, etc. (right side in Figure~\ref{fig:country-genre}) are less likely to be popular. This might be due to the paid nature of watching those videos. Due to the low volume, the videos in these genres are not salient for cultural analysis. Therefore, we exclude them in the following analysis.

\subsubsection{Measuring the relationships between video genres and objects in thumbnails}

In this section, we analyze the relationship between the object and genre in a video.
Figure~\ref{fig:genre-object} shows the distribution of trending videos among different objects and genres.
Take \textit{US} for example, Figure~\ref{fig:genre-object} shows the count numbers among different objects and genres.
It is not surprising that \{\textit{car, bus, train, truck, boat}\} have more frequently occurred in the thumbnails under \textit{Autos \& Vehicles} genre. Similarly ``Dog'' and ``Cat'' have  occurred more frequently in the \textit{Pets \& Animals} genre, while ``Person'' has occurred more frequently in the \textit{Shows} genre.

\begin{figure*}[t!]
\centering
\begin{tabular}{ccc}
    \includegraphics[width=0.3\textwidth]{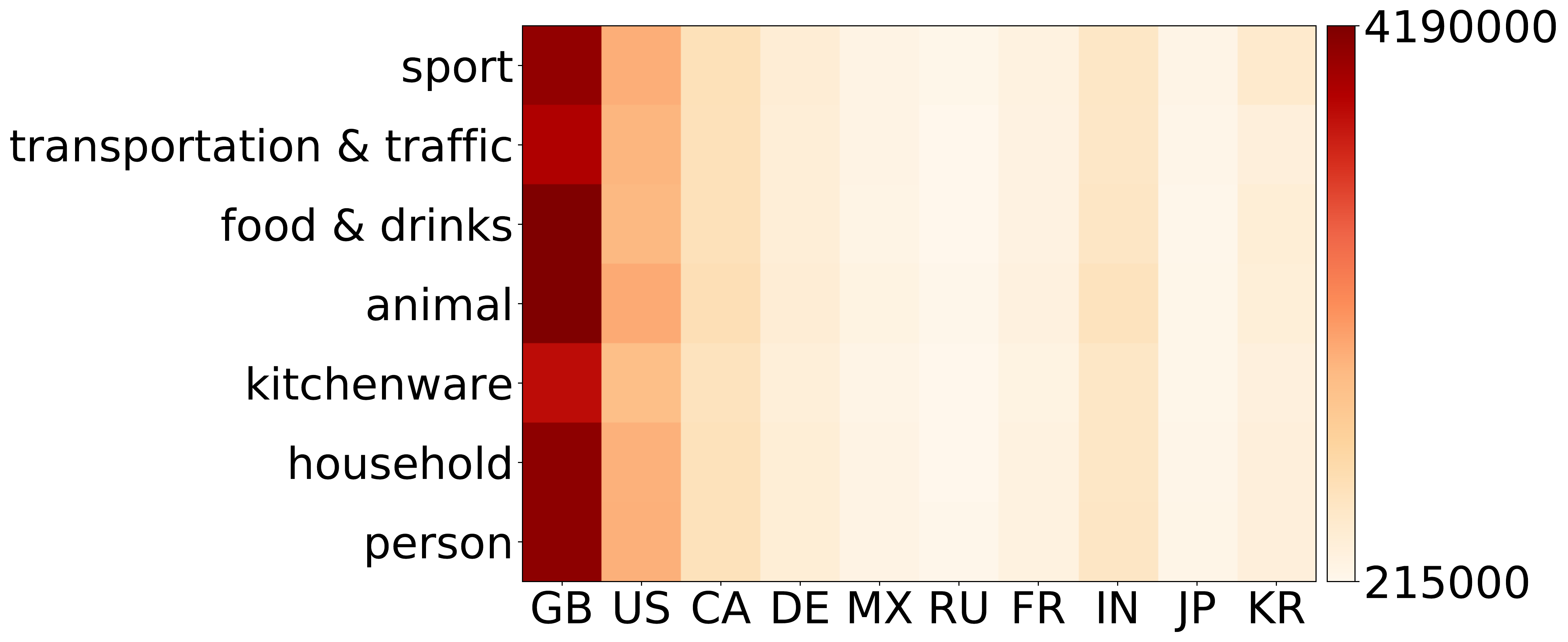} &
    \includegraphics[width=0.3\textwidth]{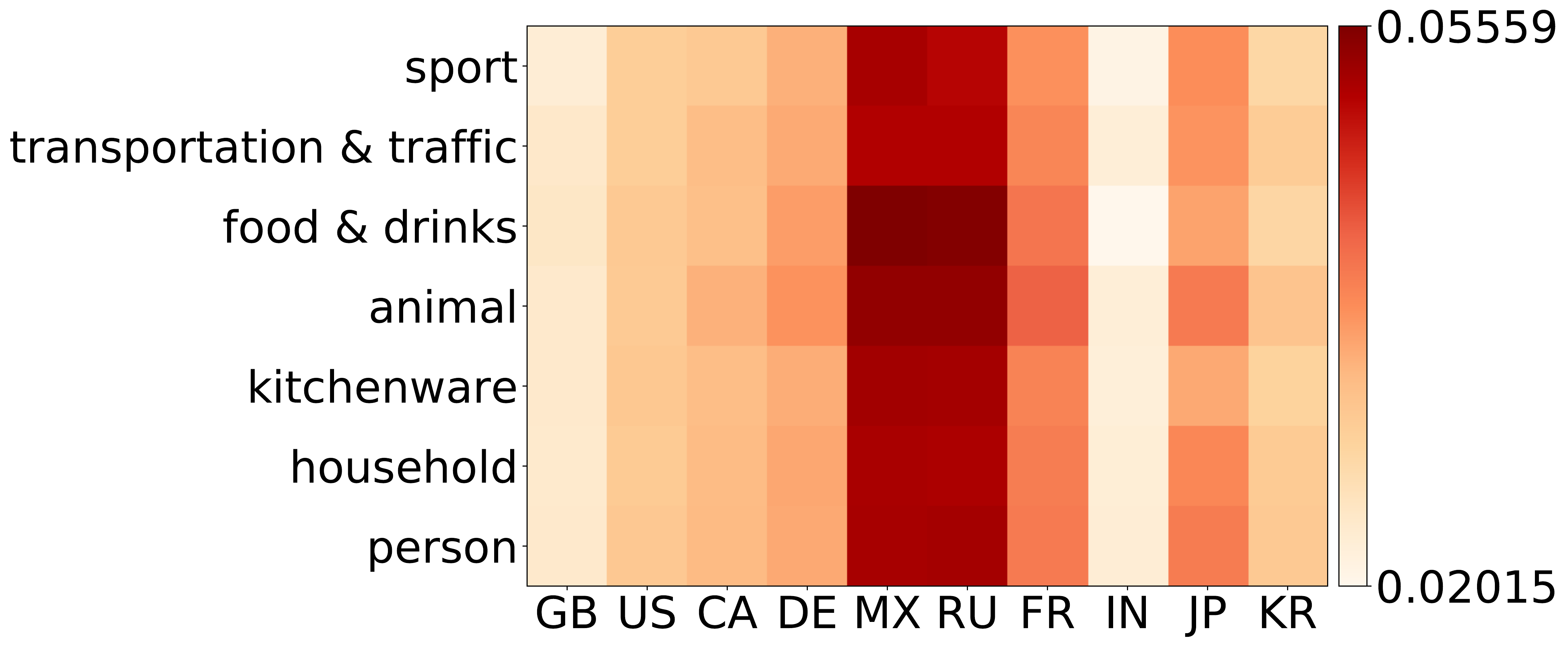} &
    \includegraphics[width=0.3\textwidth]{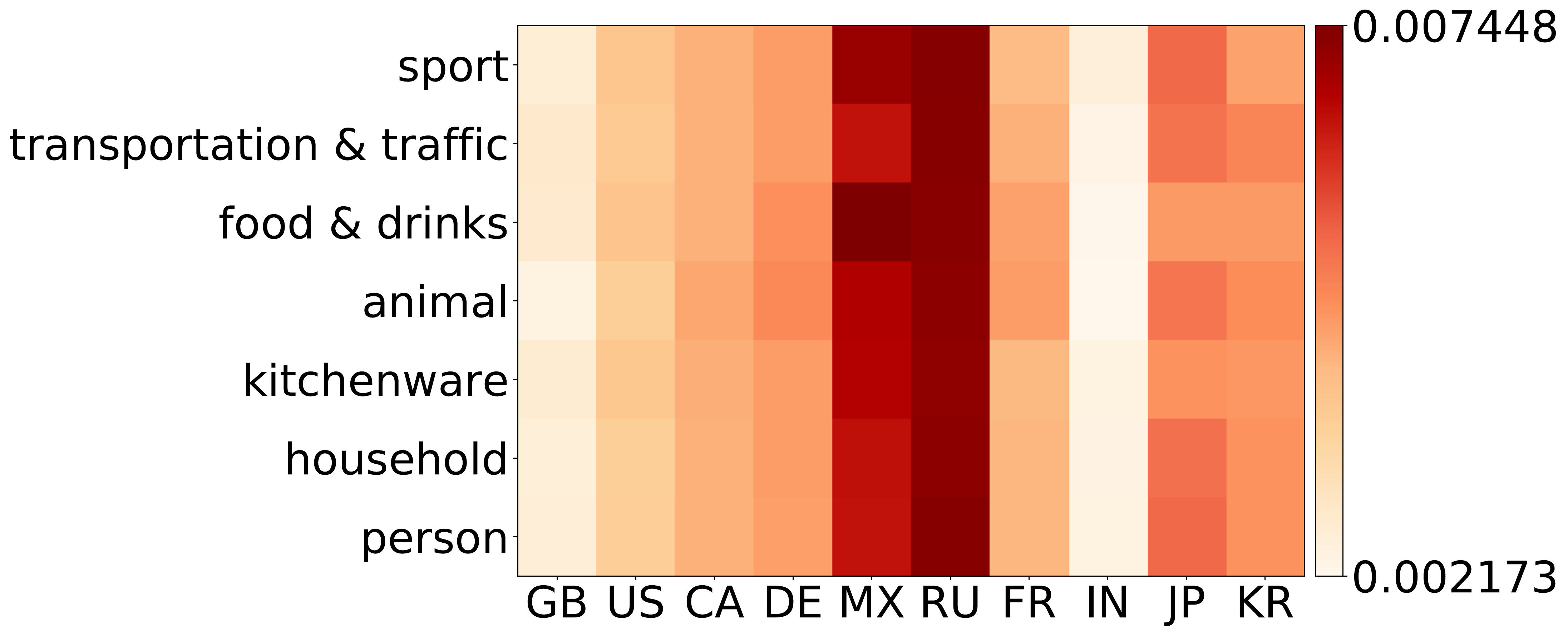} \\
    (a)  Average view & (b)  Average like per view & (c) Average comment per view \\
\end{tabular}    
    \caption{Distribution of average view, like per view and comment per view among different countries and categories.}
    \label{fig:view_like_comment}
\end{figure*}

\begin{figure*}[t!]
\centering
\begin{tabular}{ccc}
    \includegraphics[width=4.0cm]{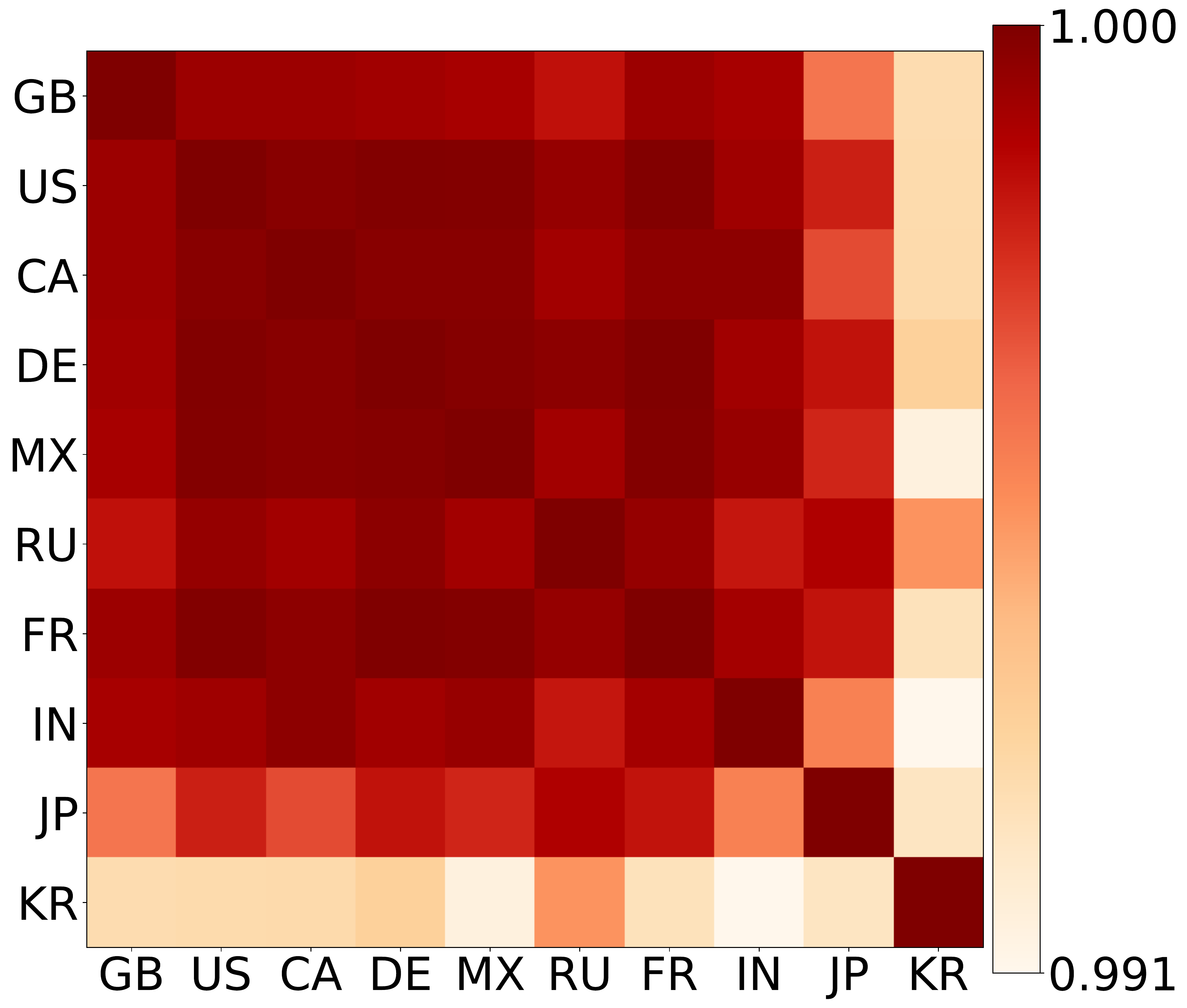} &
    \includegraphics[width=4.0cm]{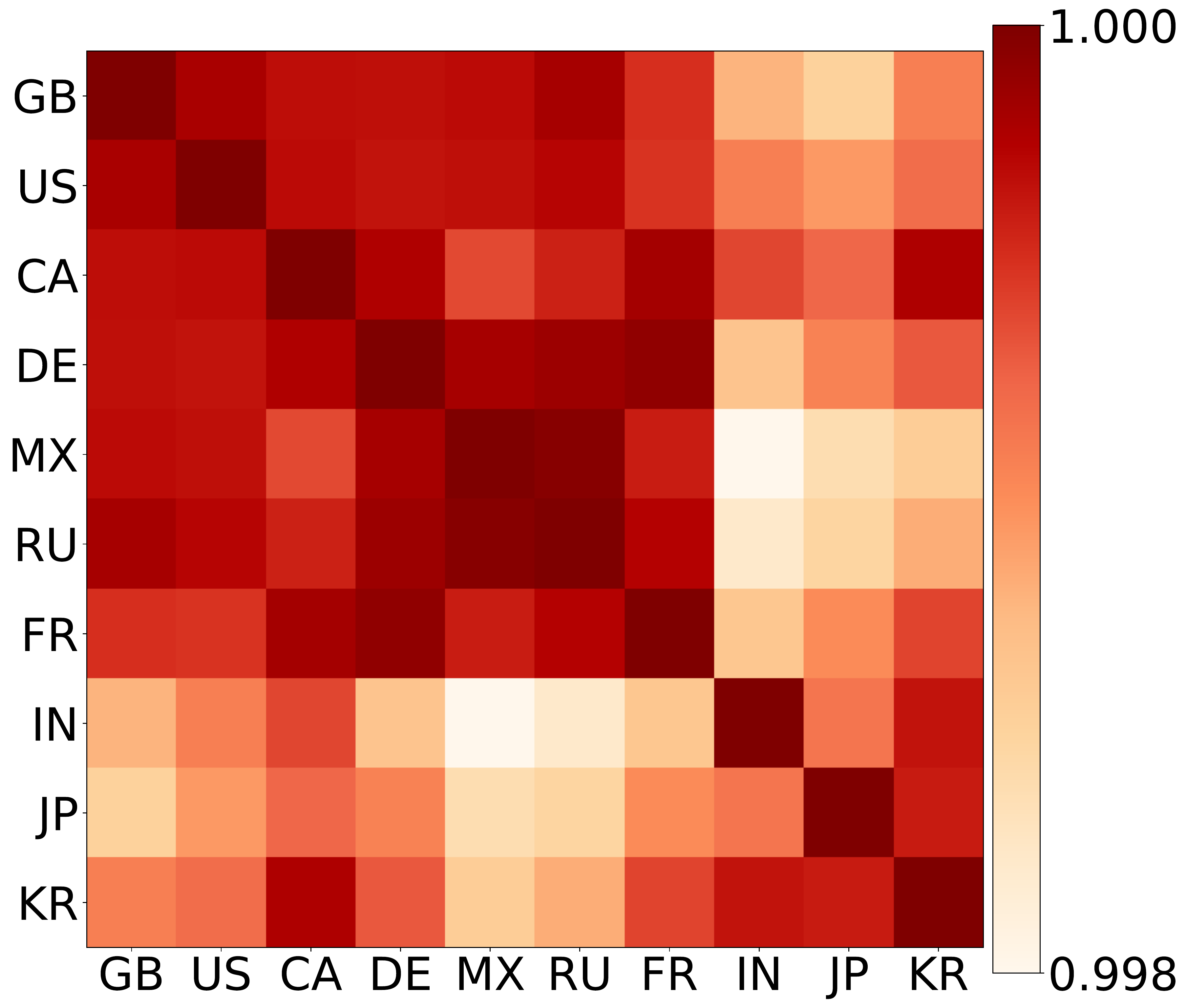} & 
    \includegraphics[width=4.0cm]{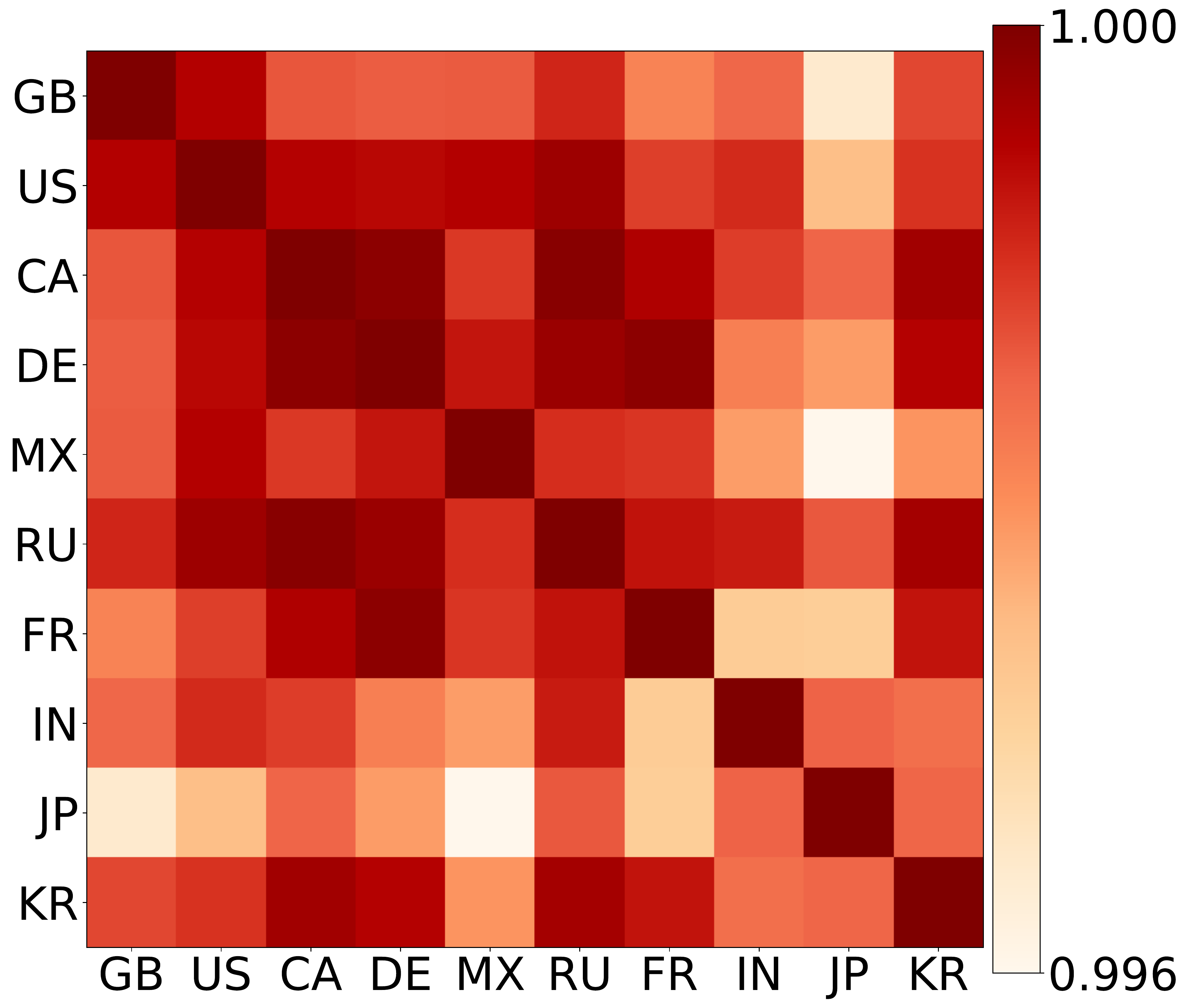} \\
    (a) Average view & (b) Average like per view & (c) Average comment per view \\
\end{tabular}
\caption{Similarity heatmap among countries in terms of view, like per view and comment per view.}
\label{fig:country_like_per_view}
\end{figure*}

\subsubsection{Determining a similarity matrix of object-genre relationship between countries}
In this section, we conduct an analysis of the similarity of object-genre relationship in different countries.
We flatten the map introduced in previous section to a single vector. This vector can present the relationship between the objects and genres for a specific country.
We then use the cosine function to compute the similarity among different countries.
Figure~\ref{fig:obj_channel_country_similarity} shows the heatmap of this similarity matrix.
From Figure~\ref{fig:obj_channel_country_similarity}, we can observe that the similarity of ``Autos \& Vehicles'' is the most diverse among all genres, (ranging from $0.884$ to $1.000$). In most plots, \textit{IN} usually shows the largest difference from other countries, which demonstrates its unique culture preference compared to other countries.
In general, we can divide the countries into three groups,  \{\textit{GB}, \textit{US}, \textit{CA}, \textit{DE}, \textit{MX}, \textit{RU}, \textit{FR}\}, \{\textit{IN}\} and \{\textit{JP}, \textit{KR}\}, where each group has common interests.



\subsection{How do YouTube users react to object distribution in thumbnails in different countries?}

In this section, we are interested in three kinds of user reactions: likes, views and comments.
The count of views can reflect how many people would like to watch the videos,
while likes per view and comments per view can usually indicate the satisfaction of users after they watch the videos.
It provides another perspective for understanding YouTube user preferences in different countries.

In this section, we first investigate how likes, views and comments are related to visual categories.
Next, we show that the similarity of such relationships between countries can reflect culture preferences.

\subsubsection{Distributions of likes, views and comments}

We first count the number of views, likes and comments across different countries and the visual categories.
Note that the thumbnails may contain multiple labels of different categories.
In this case, all the corresponding categories are counted.
The counts of likes, views and comments are then averaged by the total number of videos that contain such object labels, denoted as average likes, average views and average comments.
The average likes per view and average comments per view are computed by dividing the average likes and average comments with average views, respectively.
The heatmaps of average views, average likes per view and average comments per view are shown in Figure~\ref{fig:view_like_comment}.
From Figure~\ref{fig:view_like_comment} (a), we can observe that \textit{GB} users have higher interests in watching the trending videos.
Figure~\ref{fig:view_like_comment} (b) shows that \textit{MX} and \textit{RU} users are more willing to show their likes and opinions compared to other countries.
Comparing Figure~\ref{fig:view_like_comment} (b) with (c), it can be observed that there is a greater turnout of users pressing the ``like'' button than typing comments (range $0.020\sim 0.056$ \textit{v.s.} range $0.002\sim 0.007$).

\subsubsection{Similarity matrix between countries}

We further explore the similarity between countries in terms of likes, views and comments.
Similar to the analysis in the previous question, we use the distribution of views in a visual category in each country as its feature vector, and use the cosine similarity to measure the closeness of country pairs, as shown in 
Figure~\ref{fig:country_like_per_view}.
From Figure~\ref{fig:country_like_per_view} (a), we can observe that \textit{KR} has the most unique distribution in average views compared to others.
When looking at the average likes per view in Figure~\ref{fig:country_like_per_view} (b), there are two groups of countries \{\textit{GB, US, CA, DE, MX, RU, FR}\} and \{\textit{IN, JP, KR}\}, where each country is similar to the countries within its own group.
However, the comments per view in Figure~\ref{fig:country_like_per_view} (c) does not show such a similarity.

\section{Discussion and Conclusion}

We have shown that automated object detection of video thumbnails is an effective method to assess the similarity and difference among various cultures and countries. 
We also design a pipeline for analyzing culture preferences based on video content.
One major advantage of using the visual cues is that it do not require any translation between different countries that speak different languages, as photos and videos are truly {\it universal languages}. 
Moreover, our analyses can be universally applied to any other countries for a more comprehensive study of the different cultural characteristics of media consumption on YouTube.

Overall, our analyses suggest several important findings. First, the trending videos in YouTube demonstrate various underlying culture preferences in different countries.
The thumbnail of a video summarizes or condenses its content to a single image. 
Therefore, exploiting the visual cues in the thumbnails is an effective way to better understand culture preferences among different countries.
Previous works in cultural video analysis~\cite{park2017cultural,park2016data,baek2015relationship}  only consider metadata (video title, duration, tag, etc.).
In contrast, by exploring the underlying culture preference from the visual content, our approach allows for discovering novel insights from a more fine-grained and precise perspective.

Second, we find that even when the video genre is the same, the preference of video content may vary from country to country.
This finding is examined by analyzing the object distribution grouped by different genres.
Therefore, analyzing the visual cues from the thumbnails is also a critical mechanism for understanding the underlying culture preference in YouTube genres.

Finally, the user reactions of watching videos (likes, views, comments) vary from country to country. In some countries, the users are more willing to consume videos more often compared to those from other countries.
We also find that the users from some countries are more interested in sharing their feedbacks through comments and likes, going beyond simply viewing.
Additionally, by comparing the visual similarities, we also observe that a group of countries share similar habits in sharing videos.
Therefore, the object information in thumbnails can reveal more information on how the user behaviors are related to the cultural preferences.

\subsection{Limitations and significance of our study}

There are limitations in our approach. 
First, we assume that the thumbnails are representative enough for their corresponding videos. Such an assumption ignores the audio information, motion information and text information (if available) within the videos. These modalities are also important for analyzing the cultural differences.
For example, the trending videos coming from the ``Music'' genre can yield more insights for analyzing the cultural differences if we also consider and process the auditory information.
Likewise, using the video instead of its thumbnail image introduces several opportunities to exploit extra information such as scene understanding, activity recognition and the interaction between objects. While the  thumbnail image offers a window into the video, the video itself can possess much richer information.
Another example is that some video creators like to add texts in their thumbnails, which can act as an important cue for the user reaction through sentiment analysis and language understanding. 
Therefore, extending the analysis to other modalities in the future may generate a more comprehensive understanding of trending videos in different countries.

The dataset under study includes 10 countries chosen based on the usage rates of YouTube as well as the economic importance of their markets. While there is a diversity among these countries in terms of geography, culture and economic standings, they are still relatively closer altogether in comparisons to many other countries. Most of these countries are developed economies with significant Internet usage. Social platforms such as YouTube are influential in intercultural diffusion thus due to the similarity in high usage of Internet and economic powers, these countries may have a higher level of cultural exchange in online platforms, resulting in a convergence of cultural characteristics~\cite{Carrascosa2015QuantifyingTE}. Therefore, it is a nontrivial effort to apply stratified sampling to introduce more countries to the dataset in order to achieve a greater diversity in terms of geographical, socio-economical and educational information.

Lastly, we assume that the object detection framework make accurate prediction on the thumbnails. The errors produced by the object detection framework may influence the experimental results. In general, there are two typical errors: system errors and prediction errors.
For the system errors, since the object detection system is designed for general usage, the predicted labels may not provide enough detailed information within the thumbnails. For example, Faster-RCNN~\cite{ren2015faster} can predict whether a person exists in an image, but it cannot provide more personal information such as gestures, skin color, gender, and so on. These inputs may also be helpful for a more rigorous study.
For the prediction errors, since the model itself cannot always make the correct prediction, misclassification may influence the final results to some extent.

Due to its training scheme, the classifiers trained with the  MS-COCO datasets will detect 91 different objects under 11 categories. While MS-COCO offers a good number of objects that we come across frequently in daily lives, using a classifier with more classes and diversity can considerably improve the depth of the analysis as we can exploit more visual information beyond what a 91-class classifier allows.


\bibliography{reference}
\bibliographystyle{aaai}
\end{document}